\documentclass[final,12pt]{article}
\usepackage{hyperref}
\usepackage{amssymb,amsmath,amsthm}
\usepackage{enumerate}
\usepackage[conditional,light,first,bottomafter]{draftcopy}
\draftcopyName{DRAFT\space\today}{130}
\draftcopySetScale{65}
\usepackage[letterpaper,hmargin=3.7cm,vmargin={2.6cm,3.2cm}]{geometry}
\usepackage{setspace}
\makeatletter
\renewcommand{\section}{\@startsection%
{section}%
{1}%
{0em}%
{1.7em}%
{1.2em}%
{\normalfont\large\centering\bfseries}}
\renewcommand{\@seccntformat}[1]%
{\csname the#1\endcsname.\hspace{0.5em}}
\makeatother

\numberwithin{equation}{section}

\newtheorem{theorem}{Theorem}[section]
\newtheorem{proposition}{Proposition}[section]
\newtheorem{lemma}{Lemma}[section]
\newtheorem{corollary}{Corollary}[section]
\theoremstyle{definition}

\newtheorem*{notation}{Notation}
\newtheorem{remark}{Remark}

\newtheorem*{acknowledgments}{Acknowledgments}

\newcommand{\abs}[1]{\left|#1\right|}

\def\cprime{$'$}
\def\ocirc#1{\ifmmode\setbox0=\hbox{$#1$}\dimen0=\ht0 \advance\dimen0
  by1pt\rlap{\hbox to\wd0{\hss\raise\dimen0
  \hbox{\hskip.2em$\scriptscriptstyle\circ$}\hss}}#1\else {\accent"17 #1}\fi}

\begin{document}
\begin{titlepage}
\title{\vspace{-12mm}Discrete spectrum in a critical coupling case of
  Jacobi matrices with spectral phase transitions by uniform asymptotic analysis
\footnotetext{%
Mathematics Subject Classification(2000):
47B36, 
47A25, 
39A11, 
39A12.} 
\footnotetext{%
Keywords:
Jacobi matrices;
Uniform asymptotics;
First-order spectral phase transition;
Discrete spectrum.}
\\[2mm]}
\author{
\textbf{Serguei Naboko}\thanks{%
Author partially supported by INTAS and by
RFBR through grant 06-01-00249.}
\\
\small Department of Higher Mathematics and Mathematical Physics\\[-1.5mm]
\small Institute of Physics\\[-1.6mm]
\small St. Petersburg State University\\[-1.6mm]
\small Ulyanovskaya 1. 198904, St. Petersburg, Russia\\[-1.6mm]
\small \texttt{naboko@snoopy.phys.spbu.ru}
\\[4mm]
\textbf{Irina Pchelintseva}
\\
\small Department of Higher Mathematics and Mathematical Physics\\[-1.6mm]
\small Institute of Physics\\[-1.6mm]
\small St. Petersburg State University\\[-1.6mm]
\small Ulyanovskaya 1. 198904, St. Petersburg, Russia\\[-1.6mm]
\small \texttt{pchelintseva@yandex.ru}
\\[4mm]
\textbf{Luis O. Silva}\thanks{%
Author partially supported by CONACYT through grant P42553­F and by
PAPIIT-UNAM through grant IN-111906.}
\\
\small Departamento de M\'{e}todos Matem\'{a}ticos y Num\'{e}ricos\\[-1.6mm]
\small Instituto de Investigaciones en Matem\'aticas Aplicadas y en Sistemas\\[-1.6mm]
\small Universidad Nacional Aut\'onoma de M\'exico\\[-1.6mm]
\small C.P. 04510, M\'exico D.F.\\[-1.6mm]
\small \texttt{silva@leibniz.iimas.unam.mx}}
\date{}
\maketitle
\vspace{-4mm}
\begin{center}
\begin{minipage}{5in}
  \centerline{{\bf Abstract}} \bigskip For a two-parameter family of
  Jacobi matrices exhibiting first-order spectral phase transitions,
  we prove discreteness of the spectrum in the positive real axis when
  the parameters are in one of the transition boundaries. To this end
  we develop a method for obtaining uniform asymptotics,
  with respect to the spectral parameter, of the generalized
  eigenvectors.  Our technique can be applied to a wide range
  of Jacobi matrices.
\end{minipage}
\end{center}
\thispagestyle{empty}
\end{titlepage}
\section{Introduction}
\label{sec:intro}
In the Hilbert space $l_2(\mathbb{N})$, consider the operator $J$
whose matrix representation with respect to the canonical basis in
$l_2(\mathbb{N})$ is the Jacobi matrix
\begin{equation}
  \label{eq:jm}
  \begin{pmatrix}
  q_1 & b_1 & 0  &  0  &  \cdots \\[1mm]
  b_1 & q_2 & b_2 & 0 & \cdots \\[1mm]
  0  &  b_2  & q_3  & b_3 &  \\
  0 & 0 & b_3 & q_4 & \ddots\\
  \vdots & \vdots &  & \ddots & \ddots
  \end{pmatrix}\,,
\end{equation}
where $\{b_n\}_{n=1}^\infty\subset\mathbb{R}_+$ and
$\{q_n\}_{n=1}^\infty\subset\mathbb{R}$ (see \cite[Sec. 47]{MR1255973}
for the definition of the matrix representation of an unbounded
symmetric operator).

The spectral properties of the Jacobi operator $J$ are related with
the asymptotic behavior of the solutions of the following second order
difference system
\begin{equation}
  \label{eq:main-recurrence}
 b_{n-1}f_{n-1}(\lambda)+ q_nf_n(\lambda)+
  b_nf_{n+1}(\lambda)=
  \lambda f_n(\lambda)\,,\quad n>1\,,\quad \lambda\in\mathbb{R}\,.
\end{equation}
If a solution $\{f_n(\lambda)\}_{n=1}^\infty$ of
(\ref{eq:main-recurrence}) also satisfies
\begin{equation}
  \label{eq:initial-condition}
  q_1f_1(\lambda)+  b_1f_2(\lambda)=
  \lambda f_1(\lambda)\,
\end{equation}
then it is a generalized eigenvector of $J$. If, additionally, it
turns out that $\{f_n(\lambda)\}_{n=1}^\infty$ is in
$l_2(\mathbb{N})$, then this solution is an eigenvector of $J^*$ and
$\lambda$ is its corresponding eigenvalue. The generalized eigenvector
obtained by setting $f_1(\lambda)\equiv 1$ is the sequence of
so-called polynomials of the first kind associated to the Jacobi
operator $J$ \cite[Sec.\,2.1 Chap.\,1]{MR0184042}.

In this work we present a method for finding asymptotic expansions of
solutions of (\ref{eq:main-recurrence}) as
$n\to\infty$ which gives conditions for a uniform, with respect
to $\lambda$, estimate of the asymptotic remainder. This uniform
asymptotic method is the main result of the present work.

In the asymptotic analysis of the solutions of
(\ref{eq:main-recurrence}), the question on uniformity with respect to
$\lambda$ is particularly subtle and difficult. At the same time this
question arises in various applications and has been addressed
before. A uniform asymptotic analysis for differential equations was
carried out in \cite{MR1827087}, while the case of difference
equations was treated in \cite{MR2103377}. The results in
\cite{MR1827087} and \cite{MR2103377} were obtained by extending to
the uniform case Levinson type theorems for systems of
differential equations \cite{MR0069338,MR1006434} and difference
equations \cite{MR1002291,MR1959871}, respectively.

The uniform generalizations of Levinson type theorems for differential
and difference linear systems cannot be applied when the systems are
in the so-called ``double root'' case (see \cite{sheronova}). For
difference equations, the double root case corresponds to the product
of transfer matrices tending to a Jordan box. It turns out that
(\ref{eq:main-recurrence}) is in the double root case whenever the
corresponding Jacobi operator exhibits first-order spectral phase
transitions. An operator is said to present first-order spectral phase
transition when, by variations of parameters (phases), the spectrum of
the operator changes from absolutely continuous spectrum to discrete
spectrum or vice versa (cf. \cite{MR2312467,MR1902812,simonov-otamp}).
It is well known that the asymptotic analysis of solutions of
difference and differential equations becomes elusive and complex in
the double root case, the more so when we have an equation in the
presence of a parameter, as (\ref{eq:main-recurrence}). Our uniform
asymptotic approach has been concocted precisely for difference
equations in the double root case.

The asymptotic method proposed here is based on an asymptotic
technique due to Kelley \cite{MR1284231} and further developed by
J. Janas \cite{MR2240378}. The majorant-minorant
technique of \cite{MR1284231} allows the asymptotic analysis of
difference equations in the double root case by making use of
sequences estimating solutions of a Riccati-like difference equation
derived from (\ref{eq:main-recurrence}). We extend this approach to
consider the parametric difference equation (\ref{eq:main-recurrence})
and establish conditions for uniform, with respect to the spectral
parameter $\lambda$, asymptotic behavior of the
solutions.  We want to comment that we could not find an analogue for
our uniform asymptotic method in the theory of differential
equations. Remarkably, other methods for asymptotic analysis of
solutions of difference equations were adapted from methods in the
theory of differential equations, in particular all Levinson type
theorems.

The uniform asymptotics of solutions of linear differential systems
depending on a parameter was used in \cite{MR1849227} for the spectral
analysis of differential operators. In \cite{silva-otamp}, the uniform
asymptotic behavior of the generalized eigenvectors is pivotal in the
proof of discreteness of the spectrum for a class of Jacobi
matrices. In fact, the spectral analysis of operators is closely
related to the \emph{uniform} asymptotic analysis of the solutions of
(\ref{eq:main-recurrence}). Indeed, having a \emph{pointwise} in $\lambda$
asymptotics of the the solutions of
(\ref{eq:main-recurrence}), one can determine, by
Subordinacy theory \cite{MR915965,MR1179528}, the different parts of
the spectrum: absolutely continuous, singular continuous, and pure
point.  If one, furthermore, has \emph{uniform} estimates of the
asymptotic remainder, it is also possible to determine existence or
absence of accumulation points in intervals of the pure point part of
the spectrum. Moreover, the uniform asymptotic behavior of generalized
eigenvectors may be used to obtain estimates for the rate of accumulation
of eigenvalues at the boundaries of the pure point spectrum. We plan
to address this last topic in a forthcoming paper. It is worth noting
that uniform asymptotics of generalized eigenvectors is useful not
only for analysis of the pure point spectrum, but for other parts as
well \cite{MR1849227}.

When the asymptotic behavior of the generalized eigenvectors
exhibits certain uniformity with respect to the spectral parameter,
one may rule out accumulation points in the pure point part of
the spectrum by recurring to ideas in \cite{MR0178362,MR1137522}. This
was the approach in \cite{silva-otamp} for establishing discreteness
of the spectrum.

Surely there are methods for proving discreteness of the spectrum
which do not use uniform asymptotic analysis of the generalized
eigenvectors.  One can use for instance perturbation theory, primarily
Weyl theorem \cite{ka}, and analysis of the behavior of the operator's
quadratic form.  However these ``classical'' methods cannot be
implemented in many cases: on the one hand perturbation methods are
not so useful when the unperturbed operator has a complicated form, on
the other hand, if there are lacunae of pure point spectrum in the
continuous spectrum the quadratic form techniques become very
difficult to apply, specially when we have more than one lacuna. The
uniform asymptotic approach for ruling out accumulation points in the
pure point spectrum is local with respect to the spectral
parameter. This makes the existence of various lacunae irrelevant for
the applicability of the method.

The uniform asymptotic method proposed here is introduced by applying
it to (\ref{eq:main-recurrence}) with particular sequences
$\{b_n\}_{n=1}^\infty$ and $\{q_n\}_{n=1}^\infty$. We choose a class
of Jacobi matrices so that (\ref{eq:main-recurrence}) is in the double
root case, which implies that the uniform asymptotic analysis cannot
be carried out on the basis of uniform Levinson type theorems. In
spite of the fact the we apply our method to a particular example, it
will be clear from what follows that our approach has a general
character and may be applied in \emph{critical hyperbolic} cases
whenever the transfer matrix entries admit an asymptotic expansion in
fractional powers of $1/n$. Here, as usually in the context of
differential and difference equations, the hyperbolic case implies the
existence of increasing and decreasing solutions.

Let us briefly outline, step by step, the method for the uniform
asymptotic analysis of (\ref{eq:main-recurrence}).
\begin{enumerate}
\item \textbf{Poincar\'e type equations}. From the difference equation
  (\ref{eq:main-recurrence}), we obtain two Poincar\'e type equations
  with smooth, with respect to $n$, coefficients (see
  (\ref{eq:birkhoff-adams-eq-x}) and (\ref{eq:birkhoff-adams-eq-y})
  below). This step is straightforward, however we should remark that,
  in our case, the requirement for the Poincar\'e coefficients to be
  smooth yields a system of two second-order difference equations
  (cf. \cite{MR1911189,simonov-otamp}). The asymptotic analysis of
  (\ref{eq:main-recurrence}) is carried out through the analysis of
  \emph{one} of the obtained Poincar\'e type equations since the
  asymptotic behavior of the solutions of one equation determines
  straightforwardly the asymptotic behavior of the other.
\item \textbf{Riccati difference equation}.  We derive a Riccati
  difference equation (see (\ref{eq:kelly-eq}) below) from one of the
  Poincar\'e type equations. This is done by a change of variable as
  in \cite{MR1284231} which is a particular realization of the
  so-called Riccati transformations
  (cf. \cite[Sec. 7.2]{MR2128146}). These transformations are well
  known in the continuous case (cf. \cite[Sec.\,6.1]{olver}).
\item \textbf{Formal asymptotic expansion}. We give an iterative
  procedure that provides the heuristic for
  obtaining a formal asymptotic expansion of two solutions
  $\{X_n^\pm(\lambda)\}$ of (\ref{eq:kelly-eq}). This step is an
  important component of the method since the next steps are based on
  the formal asymptotics found here.
\item \textbf{Uniform majorant and minorant sequences}. By the
  introduction of a new parameter in one term of the formal asymptotic
  expansions of $\{X_n^\pm(\lambda)\}$, we explicitly
  construct new parametrized sequences. Then we give conditions on the
  parameters so that the parametrized sequences serve as majorant and
  minorant sequences satisfying the hypothesis of
  Propositions~\ref{thm:kelley1} and \ref{thm:kelley2}. All this works
  under the requirement that the terms of the asymptotic expansion of
  the Poincar\'{e} coefficients, up to a certain order, are
  ``differentiable'' with respect to $n$. Thus, in this step we also
  obtain how precise our calculations for the asymptotic expansion of
  the Poincar\'{e} coefficients should be.
\item \textbf{Uniform asymptotics of solutions of the Riccati
    equation}. Using the majorant and minorant sequences found in the
  previous step, we obtain uniform estimating sequences for
  $\{X_n^\pm(\lambda)\}$  by applying the
  straightforward generalizations of Kelley's theorems, namely
  Propositions \ref{thm:kelley1} and \ref{thm:kelley2}. The uniform
  estimating sequences allow us to prove the asymptotic formulae for
  $\{X_n^\pm(\lambda)\}$  with a uniform
  estimate of the asymptotic remainder.
\item \textbf{Uniform asymptotics of the solutions of the generalized
    eigenvectors}. From the uniform asymptotic formulae for
  $\{X_n^\pm(\lambda)\}$, we obtain the uniform asymptotic behavior of
  a pair of solutions of the Poincar\'e equation
  (\ref{eq:birkhoff-adams-eq-x}). With this information we found the
  uniform asymptotic expansion of two linearly independent solutions
  of (\ref{eq:main-recurrence}).
\end{enumerate}

Having the asymptotics of the solutions of (\ref{eq:main-recurrence}),
we make use of Subordinacy theory to prove pure point spectrum on
$\mathbb{R}_+$. Then, on the basis of the \emph{uniform} asymptotics
of the generalized eigenvectors, and their smoothness with respect to
$\lambda$, we show that there are no accumulation points in the pure
point part of the spectrum, excluding its boundaries. This fact is
proven by the technique used in \cite{silva-otamp}.

We stress the fact that, for the class of Jacobi matrices discussed
here, the classical methods for proving discreteness of the spectrum
mentioned above have proven somehow difficult to apply because of the
form of the unperturbed operator. It is also worth remarking that the
spectral analysis of operators associated with these matrices may be
relevant in itself. Our conclusions here shed light on the spectral
properties of a two-parameter family of Jacobi operators exhibiting a
first-order spectral phase transition. We prove discreteness of the
spectrum in the positive real axis when the parameters are in one of
the transition boundaries.

The paper is organized as follows. In Section~\ref{sec:preliminaries}
we lay down the notation, introduce the two-parameter family of Jacobi
operators, and present some preparatory facts. Here we take the first
two steps of the outline above.
Section~\ref{sec:formal-uniform-asymptotics} presents step 3.  In
Section~\ref{sec:kelley estimates} we carry out steps 4 and
5. Section~\ref{sec:generalized-eigenvectors} contains step 6. In
Section~\ref{sec:discrete-spectrum} we give the spectral
characterization of the class of Jacobi operators under
consideration. Finally, the details of calculations for the asymptotic
of some concrete sequences can be found in the Appendix.

\section{Preliminaries}
\label{sec:preliminaries}
In this section we introduce the notation, a two-parameter family of
Jacobi operators, and some preliminary facts. In particular, we
present the main difference equations whose asymptotic analysis is
carried out in subsequent sections.
\begin{notation}
\begin{enumerate}
\item Throughout this paper a sequence of numbers, depending on a real
  parameter $\lambda$ and enumerated from some $N\in\mathbb{N}$, will
  be denoted by $\{g_n(\lambda)\}_{n=N}^\infty$. In general, $N$ plays
  no r\^ole in a discussion focused on the asymptotic behavior of
  $\{g_n(\lambda)\}_{n=N}^\infty$ as $n\to\infty$. Nevertheless, since
  our goal are asymptotic expansions of
  $\{g_n(\lambda)\}_{n=N}^\infty$, uniform with respect to $\lambda$,
  we should watch over $N$ and its possible dependence on
  $\lambda$. Having said this, we shall sometimes write
  $\{g_n(\lambda)\}$  instead of
  $\{g_n(\lambda)\}_{n=N}^\infty$, whenever no confusion is likely to
  arise.\label{notation1}
\item Along with the standard notation of number sets, $\mathbb{N}$,
  $\mathbb{R}$, we use $\mathbb{R}_+$ to denote the set of real numbers
  greater than zero.

\item Consider a set $I\subset\mathbb{R}$, a sequence $\{g_n(\lambda)\}$,
  depending on a real parameter $\lambda$, and
  a sequence $\{h_n\}$ of real numbers. We shall say that
\begin{equation*}
  g_n(\lambda)=\widetilde{O}_I(h_n)\quad\text{ as }
  n\to\infty\,
\end{equation*}
if there exists a constant $C>0$ and $N\in\mathbb{N}$ such
that
\begin{equation*}
  \sup_{\lambda\in I}\abs{g_n(\lambda)}<C\abs{h_n}\,,
  \qquad n>N\,.
\end{equation*}\label{notation2}
\item Clearly, in the previous item, the constants $N$ and $C$ are
  supposed not to depend on $\lambda\in I$. There and \emph{in the sequel
  auxiliary constants are assumed to be independent of $\lambda$},
  unless we indicate the dependency explicitly.\label{notation3}
\item Let $I\subset\mathbb{R}$, $\{g_n(\lambda)\}$ be a sequence
  depending on a real parameter $\lambda$, and $\{h_n\}$ be a sequence of
  real numbers. If for any $\epsilon>0$ there exists $N\in\mathbb{N}$
  such that
  \begin{equation*}
    \frac{\sup_{\lambda\in I}\abs{g_n(\lambda)}}{\abs{h_n}}<\epsilon\,,
  \qquad n>N\,,
  \end{equation*}
then we say that
\begin{equation*}
  g_n(\lambda)=\widetilde{o}_I(h_n)\quad\text{ as }
  n\to\infty\,.
\end{equation*}\label{notation4}
\end{enumerate}
\end{notation}

Let us now introduce the class of operators for which we provide a
uniform asymptotic analysis of the generalized eigenvectors.

In the Hilbert space $l_2(\mathbb{N})$, we define the Jacobi operator
$J$ as the one whose matrix representation with respect to the
canonical basis in $l_2(\mathbb{N})$ is (\ref{eq:jm}) (we refer to
\cite[Sec.\,47]{MR1255973} for a discussion on matrix representation
of unbounded symmetric operators), where the sequences
$\{b_n\}_{n\in\mathbb{N}}$ and $\{q_n\}_{n\in\mathbb{N}}$ are defined by
\begin{equation}
  \label{eq:weights-diagonal}
  \begin{split}
  b_n := n^{\alpha}c_n\,,&\qquad q_n:=n^{\alpha}
  \qquad  n\in\mathbb{N}\,,\\
  c_{2n-1}=c_1\,,&\qquad c_{2n}=c_2\,,
  \qquad c_1,c_2\in\mathbb{R}\setminus \{0\}\,,
\end{split}
\end{equation}
with
\begin{equation}
  \label{eq:alpha-def}
 \alpha\in (1/3,1/2)\,.
\end{equation}
This particular choice of $\alpha$ is made for simplifying some
calculations. We could have take $0<\alpha<1$, but then the derivations
of some asymptotic formulae would have been hindered by algebraic
technicalities by no means important for our considerations.

Allowing $c_1,c_2$ to vary through $\mathbb{R}\setminus\{0\}$, one
obtains a two-parameter family of Jacobi operators $J=J(c_1,c_2)$. Due
to the Carleman criterion \cite[Thm.\,1.3 Chap.\,7]{MR0222718},
(\ref{eq:weights-diagonal}) implies that $J$ is self-adjoint for any
$c_1,c_2\in\mathbb{R}\setminus\{0\}$.

As it was shown in \cite{simonov-otamp} for the case $\alpha=1$, one
can immediately conclude, on the basis of results on periodically
modulated Jacobi matrices \cite{MR1911189}, that $J$ exhibits a
first-order spectral phase transition for all $\alpha\in(0,1]$. There is a
region, $\abs{\frac{c_1^2+c_2^2-1}{c_1c_2}}<2$, where the spectrum of
$J$, henceforth denoted by $\sigma(J)$, is purely absolutely continuous
and it covers the whole real line. In the region
$\abs{\frac{c_1^2+c_2^2-1}{c_1c_2}}>2$ the spectrum is discrete, that
is, $\sigma(J)=\sigma_{disc}(J)$ (see
Figure~\ref{fig1}).

\unitlength=0.45mm \newcounter{N}
\begin{figure}[h]
\centering
\begin{picture}(145,110)
\put(50,0){\vector(0,1){106}} \put(0,50){\vector(1,0){106}}
\put(25,0){\line(1,1){75}}
\put(0,25){\line(1,1){75}}
\put(0,75){\line(1,-1){75}}
\put(25,100){\line(1,-1){75}}
\put(45,44){\footnotesize 0}
\put(47,75){\line(1,0){5}}
\put(42,72){\footnotesize 1}
\put(47,108){\small $c_2$}
\put(75,47){\line(0,1){5}}
\put(73,40){\footnotesize 1}
\put(108,47){\small $c_1$}
\linethickness{0.1pt}
\setcounter{N}{50}
\multiput(0,25)(2,2){13}
{\line(0,1){\value{N}}
\addtocounter{N}{-4}}
\setcounter{N}{50}
\multiput(50,25)(2,2){13}
{\line(0,1){\value{N}}
\addtocounter{N}{-4}}
\setcounter{N}{0}
\multiput(25,50)(2,-2){13}
{\line(0,1){\value{N}}
\addtocounter{N}{4}}
\setcounter{N}{0}
\multiput(75,50)(2,-2){13}
{\line(0,1){\value{N}}
\addtocounter{N}{4}}
\setcounter{N}{0}
\multiput(25,100)(2,-2){13}
{\line(0,1){\value{N}}
\addtocounter{N}{2}}
\setcounter{N}{0}
\multiput(25,0)(2,2){13}
{\line(0,-1){\value{N}}
\addtocounter{N}{2}}
\setcounter{N}{25}
\multiput(50,75)(2,2){13}
{\line(0,1){\value{N}}
\addtocounter{N}{-2}}
\setcounter{N}{25}
\multiput(50,25)(2,-2){13}
{\line(0,-1){\value{N}}
\addtocounter{N}{-2}}
\put(125,78){\framebox(18,6){}}
\put(125,23){\framebox(18,6){}}
\multiput(127,78)(2,0){8}
{\line(0,1){6}}
\put(120,65){\footnotesize $\sigma(J)=\sigma_{\rm
    disc}(J)$}
\put(120,10){\footnotesize $\sigma(J)=\sigma_{\rm a.c}(J)$}
\end{picture}
\caption{Phase space} \label{fig1}
\end{figure}
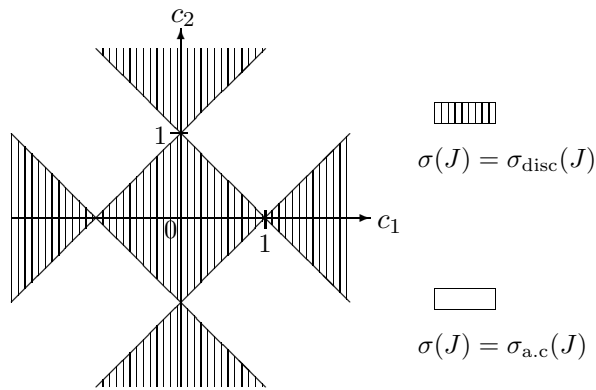

The symmetry of Figure~\ref{fig1} obeys the existence of unitary
mappings that transform $J$ symmetrically for one quadrant of the
plane $(c_1,c_2)$ to another
(see \cite[Lem.\,1.6]{MR1711536}). Thus, the study of the interplay
between the spectral properties of $J$ and the coefficients $c_1,c_2$
can be constrained to the case
\begin{equation}
  \label{eq:c1-c2-positivity}
  c_1,\,c_2>0\,.
\end{equation}

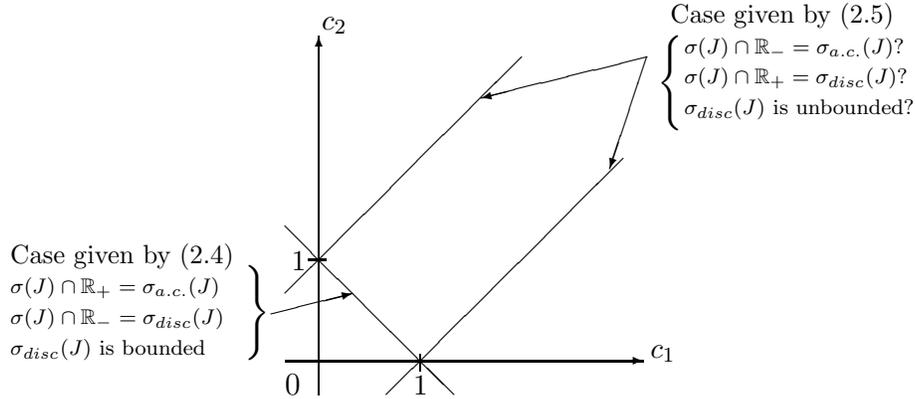
\begin{figure}[h]
\centering
\begin{picture}(100,110)
\put(10,0){\vector(0,1){106}} \put(0,10){\vector(1,0){106}}
\put(0,50){\line(1,-1){50}}
\put(30,0){\line(1,1){70}}
\put(0,30){\line(1,1){70}}
\put(0,0){0}
\put(7,40){\line(1,0){5}}
\put(2,37){\small 1}
\put(11,108){\small $c_2$}
\put(40,7){\line(0,1){5}}
\put(38,0){\small 1}
\put(108,11){\small $c_1$}
\put(-4,24){\vector(4,1){24}}
\put(-12,22){$\Biggr\}$}
\put(-81,39){\footnotesize Case given by (\ref{easy-case})}
\put(-81,30){\scriptsize $\sigma(J)\cap\mathbb{R}_+=\sigma_{a.c.}(J)$}
\put(-81,21){\scriptsize $\sigma(J)\cap\mathbb{R}_-=\sigma_{disc}(J)$}
\put(-81,12){\scriptsize $\sigma_{disc}(J)$ is bounded}
\put(107,100){\vector(-4,-1){49}}
\put(107,100){\vector(-1,-3){11}}
\put(114,110){\footnotesize Case given by (\ref{complicated-case})}
\put(110,90){$\Biggl\{$}
\put(118,101){\scriptsize $\sigma(J)\cap\mathbb{R}_-=\sigma_{a.c.}(J)$?}
\put(118,92){\scriptsize $\sigma(J)\cap\mathbb{R}_+=\sigma_{disc}(J)$?}
\put(118,83){\scriptsize $\sigma_{disc}(J)$ is unbounded?}
\end{picture}
\caption{Transition edges}\label{fig.2}
\end{figure}

The transition boundary $\abs{\frac{c_1^2+c_2^2-1}{c_1c_2}}=2$ may be
split into two different cases by the conditions
\begin{align}
  c_1+c_2&=1\label{easy-case}\\
 \abs{c_1-c_2}&=1\label{complicated-case}\,.
\end{align}
The spectral properties of $J$ in the case (\ref{easy-case}) can be
revealed by means of the asymptotic techniques used in
\cite{simonov-otamp} and Subordinacy theory
\cite{MR915965,MR1179528}. The conclusion is that, for
$\alpha\in(0,1)$ and all $c_1,c_2$
satisfying (\ref{eq:c1-c2-positivity}) and (\ref{easy-case}), the
spectrum of $J$ is absolutely continuous on $\mathbb{R}_+$ and
discrete on $\mathbb{R}_-$ (see Figure~\ref{fig.2}). One excludes the
possibility of accumulation points in the pure point part of the
spectrum by repeating the reasoning of \cite{simonov-otamp} which
relies on estimates of the quadratic form of $J$ and a theorem due to
Glazman \cite[Sec. 3 Thm. 6]{MR0185471}. We draw the reader's attention to the fact
that in the case $\alpha=1$, with $c_1,c_2$ such that
(\ref{eq:c1-c2-positivity}) and (\ref{easy-case}) holds, the spectrum
of $J$ is purely absolutely continuous on $(\frac{1}{2},+\infty)$  and
discrete on $(0, \frac{1}{2})$ \cite[Cor.\,3.4]{simonov-otamp}.

The asymptotic analysis of (\ref{eq:main-recurrence}) will be based on
a modification of Kelley's method for computing asymptotic
approximations to solutions of difference equations. The starting
point is the following elementary assertion by which we derive from
(\ref{eq:main-recurrence}) Poincar\'{e} type equations
(cf. \cite[Sec. 8.2]{MR2128146} and the original works
\cite{perron,poincare}) with smooth in $n$ coefficients. An analogous
assertion can be found in \cite{simonov-otamp}.
\begin{lemma}
  \label{lem:PP-equation}
  Let $\{b_n\}_{n=1}^\infty\subset\mathbb{R}_+$ and
$\{q_n\}_{n=1}^\infty\subset\mathbb{R}$.
  Assume that there is a number
  $N\in\mathbb{N}$ and a set $I\subset\mathbb{R}$ such that
  $q_{2n}\not\in I$ for all $n> N$. For $n\ge N$ and
  $\lambda\in I$, define
\begin{equation}
  \label{eq:f-g-definition}
  \begin{split}
  F_n(\lambda)&:=\frac{(q_{2n+2}-\lambda)b_{2n}^2}
 {(q_{2n}-\lambda)b_{2n+1}b_{2n+2}}-
\frac{(q_{2n+1}-\lambda)(q_{2n+2}-\lambda)}{b_{2n+1}b_{2n+2}}
+ \frac{b_{2n+1}}{b_{2n+2}}\,,\\
 G_n(\lambda)&:=\frac{(q_{2n+2}-\lambda)b_{2n-1}b_{2n}}
 {(q_{2n}-\lambda)b_{2n+1}b_{2n+2}}\,,
\end{split}
\end{equation}
Then, for $\lambda\in I$, the
  sequence $\{f_n(\lambda)\}_{n=2N}^\infty$ satisfies
  (\ref{eq:main-recurrence}) for $n> 2N$ iff the
  sequences $\{x_n(\lambda)\}_{n=N}^\infty$ and
  $\{y_n(\lambda)\}_{n=N}^\infty$ given by
\begin{equation}
  \label{eq:x-y-sequences}
  x_n(\lambda):=f_{2n+1}(\lambda)\,,\qquad
  y_n(\lambda):=f_{2n}(\lambda)\,,\quad \lambda\in I
\end{equation}
satisfy the following Poincar\'{e} type equations
\begin{align}
\label{eq:birkhoff-adams-eq-x}
  x_{n+1}(\lambda)+F_n(\lambda)x_n(\lambda)+G_n(\lambda)x_{n-1}(\lambda)&=0\,,
\qquad \lambda\in I\,,\  n> N\,,\\
  \label{eq:birkhoff-adams-eq-y}
   y_{n+1}(\lambda)+ F_{n-\frac{1}{2}}(\lambda)y_n(\lambda)+
G_{n-\frac{1}{2}}(\lambda)y_{n-1}(\lambda)&=0\,,
\qquad \lambda\in I\,,\ n> N\,.
\end{align}
and the conditions
\begin{align}
 b_{2n-2}y_{n-1}(\lambda)+q_{2n-1}x_{n-1}(\lambda)+b_{2n-1}y_{n}(\lambda)&=\lambda
 x_{n-1}(\lambda)\label{eq:null-eq}
\\
  b_{2n-1}x_{n-1}(\lambda)+q_{2n}y_n(\lambda)+b_{2n}x_n(\lambda)&=\lambda
  y_n(\lambda)\,,\label{eq:all_determined_first_eq}
\end{align}
for all $n>N$.
\end{lemma}
\begin{proof}
  Write down (\ref{eq:main-recurrence}) with $n=2n,2n+1,2n+2$. From
  the equations with $n=2n+1,2n+2$ express $f_{2n}(\lambda)$ in terms of
  $f_{2n+1}(\lambda)$ and $f_{2n+3}(\lambda)$. Substitute this into the equation with
  $n=2n$. It is now straightforward to verify that
  $\{x_n(\lambda)\}$ satisfies
  (\ref{eq:birkhoff-adams-eq-x})
Analogously, one proves that $\{y_n(\lambda)\}$ satisfies
(\ref{eq:birkhoff-adams-eq-y}) by considering
(\ref{eq:main-recurrence}) for $n=2n-1,2n,2n+1$, then  expressing $f_{2n-1}(\lambda)$
through $f_{2n}(\lambda)$ and $f_{2n+2}(\lambda)$ and substituting
this into the equation with $n=2n-1$.
\end{proof}
The sequences $\{F_n(\lambda)\},\,\{G_n(\lambda)\}$ will be called the Poincar\'{e}
coefficients.
\begin{remark}
  \label{rem:N}
  Let $I$ be any bounded interval of $\mathbb{R}_+$. If the sequences
  $\{b_n\}$ and $\{q_n\}$ are given by (\ref{eq:weights-diagonal})
  then the conditions of Lemma~\ref{lem:PP-equation} are satisfied
  once we choose a natural number $N\ge\frac{(\sup I)^{1/\alpha}}{2}$.
\end{remark}
\begin{remark}
  \label{rem:poincare}
  Note that the Poincar\'e coefficients defined in
  (\ref{eq:f-g-definition}) are smooth in $n$ when the
  sequences $\{b_n\}$ and $\{q_n\}$ are given by
  (\ref{eq:weights-diagonal}) and (\ref{complicated-case}). Of course,
  one could have obtained directly from (\ref{eq:main-recurrence}) a
  single Poincar\'{e} type equation, but, for our choice of $\{b_n\}$
  and $\{q_n\}$, the corresponding Poincar\'e coefficients would not
  have been smooth in $n$.
\end{remark}
\begin{remark}
\label{rem:rem}
  Any solution $\{f_n(\lambda)\}_{n=1}^\infty$ of the three term
  recurrence equation
  (\ref{eq:main-recurrence}) is uniquely determined by any two consecutive
  elements of the sequence $\{f_n(\lambda)\}_{n=1}^\infty$ for each
  $\lambda$. The
same is true for solutions $\{x_n(\lambda)\}_{n=N}^\infty$ and
$\{y_n(\lambda)\}_{n=N}^\infty$ of (\ref{eq:birkhoff-adams-eq-x}) and
(\ref{eq:birkhoff-adams-eq-y}), respectively. Moreover, assuming that
$N\in\mathbb{N}$ and $I\subset\mathbb{R}$ satisfy the conditions of
Lemma~\ref{lem:PP-equation}, it is easy to show from
(\ref{eq:all_determined_first_eq}) that two consecutive elements of
$\{x_n(\lambda)=f_{2n+1}\}_{n=N}^\infty$ uniquely determine any
solution $\{f_n(\lambda)\}_{n=2N+1}^\infty$ of
(\ref{eq:main-recurrence}) for $n>2N+1$.
\end{remark}

The following straightforward assertion shows how to obtain a Riccati
difference equation from a Poincar\'{e} type equation.
\begin{lemma}
  \label{lem:Kelley-equation}
  Consider (\ref{eq:birkhoff-adams-eq-x}) with arbitrary coefficients
  $\{F_n(\lambda)\},\,\{G_n(\lambda)\}$ depending on a parameter
  $\lambda\in I$.  Let us suppose that one can find a natural number
  $N$ such that
 \begin{equation*}
   \begin{split}
   x_n(\lambda)&\ne 0\,,\qquad \lambda\in I\,,\quad n>N-1\\
   F_n(\lambda)&\ne 0\,,\qquad \lambda\in I\,,\quad n\ge N-1\,.
   \end{split}
 \end{equation*}
For $n>N$ and $\lambda\in I$ define new coefficients
\begin{equation}
  \label{eq:beta-x-def}
  \beta_n(\lambda):=\frac{4G_n(\lambda)}{F_n(\lambda)F_{n-1}(\lambda)}-1\,,
\end{equation}
and a new variable
\begin{equation}
  \label{eq:x-definition}
  X_n(\lambda):=\frac{-2x_{n+1}(\lambda)}{F_n(\lambda)x_n(\lambda)}-1\,.
\end{equation}
Then, $\{x_n(\lambda)\}_{n=N}^\infty$ satisfies (\ref{eq:birkhoff-adams-eq-x}) for
$n>N$ and $\lambda\in I$ iff the new sequence $\{X_n(\lambda)\}_{n=N}^\infty$
satisfies the Riccati difference equation
\begin{equation}
  \label{eq:kelly-eq}
  X_n(\lambda)=(1+\beta_n(\lambda))
\frac{X_{n-1}(\lambda)}{X_{n-1}(\lambda)+1} -\beta_n(\lambda)\,,
\qquad \lambda\in I\,,\quad n>N\,.
\end{equation}
\end{lemma}
\begin{proof}
Consider the sequence $\{\xi_n(\lambda)\}_{n=N}^\infty$,
whose elements are given by
\begin{equation}
  \label{eq:definition-xi}
  \xi_n(\lambda):=x_n(\lambda)\prod_{k=N-1}^{n-1}\frac{-2}{F_k(\lambda)}\,,
\qquad \lambda\in I
\end{equation}
Substituting this expression into (\ref{eq:birkhoff-adams-eq-x}) for
$n> N$ (cf. \cite[Sec. 8.5]{MR2128146}) , one arrives at
\begin{equation}
\label{eq:aux-sys}
  \xi_{n+1}(\lambda)-2\xi_n(\lambda) + \frac{4G_n(\lambda)}
   {F_n(\lambda)F_{n-1}(\lambda)}\xi_{n-1}(\lambda) = 0\,,
\quad \lambda\in I\,,\  n> N\,.
\end{equation}
Taking into account that
$X_n(\lambda)=\frac{\xi_{n+1}(\lambda)}{\xi_n(\lambda)}-1$, one easily
obtains that (\ref{eq:aux-sys}) is equivalent to (\ref{eq:kelly-eq})
(cf. \cite{MR1284231}).
\end{proof}
\begin{remark}
  \label{rem:riccati}
  Let $\{b_n\}$, $\{q_n\}$ be defined by
  (\ref{eq:weights-diagonal}). On the basis of Remark~\ref{rem:N} one
  may consider (\ref{eq:birkhoff-adams-eq-x}), with $\{F_n(\lambda)\}$
  and $\{G_n(\lambda)\}$ given by (\ref{eq:f-g-definition}), for
  $N_1>\frac{(\sup I)^{1/\alpha}}{2}$ and $I$ being a bounded interval
  of $\mathbb{R}_+$. If (\ref{complicated-case}) holds and there is
  $N_2\ge N_1$ such that $x_n(\lambda)\ne 0$ for all $n\ge N_2$ and
  $\lambda\in I$, then the conditions of
  Lemma~\ref{lem:Kelley-equation} are satisfied for a certain
  $N\ge N_2$. This follows straightforwardly from the uniform asymptotic
  formula (\ref{eq:F-n-expansion}) for $\{F_n(\lambda)\}$ as $n\to\infty$.
\end{remark}
We shall use the following elementary but important results. They are
the uniform counterparts of \cite[Thm.\,1]{MR1284231} and
\cite[Thm.\,2]{MR1284231}.
\begin{proposition}
  \label{thm:kelley1}
  Let $I$ be a subset of $\mathbb{R}$.  Suppose that one can find
  $N\in\mathbb{N}$ and real sequences $\{v_n(\lambda)\}$ and
  $\{w_n(\lambda)\}$ such that
  \begin{align}
    \inf_{\lambda\in I}v_n(\lambda)&>
    -1\,,\qquad n\ge N\label{eq:kelley-3}\\
 w_{N}(\lambda)&\ge v_{N}(\lambda)\,,\qquad \lambda\in I\,,\label{eq:kelley-4}
  \end{align}
and, for all $n>N$,
 \begin{align}
   0 &\le 1+\beta_n(\lambda)\,,\quad\lambda\in I\label{eq:kelley-0}\\
    v_n(\lambda)&\le\frac{(1+\beta_n(\lambda))v_{n-1}(\lambda)}{1+v_{n-1}(\lambda)}-
\beta_n(\lambda)\,,\quad\lambda\in I
    \label{eq:kelley-1}\\
w_n(\lambda)&\ge\frac{(1+\beta_n(\lambda))w_{n-1}(\lambda)}{1+w_{n-1}(\lambda)}
-\beta_n(\lambda)\,,\quad\lambda\in I\,.
    \label{eq:kelley-2}
  \end{align}
 Suppose that $\{X_n(\lambda)\}_{n=N}^\infty$ satisfies
  (\ref{eq:kelly-eq}) for $n>N$ and $\lambda\in I$. Moreover, let
  $X_{N}(\lambda)\in
  [v_{N}(\lambda),w_{N}(\lambda)]$ for all
  $\lambda\in I$, then
 \begin{equation*}
    v_n(\lambda)\le X_n(\lambda)\le w_n(\lambda)\,,\qquad  n\ge
    N\,,\quad \lambda\in I\,.
  \end{equation*}
\end{proposition}
\begin{proof}
  The statement follows straightforwardly from the proof of
  \cite[Thm.\,1]{MR1284231}. Here we repeat almost verbatim the proof of
  \cite[Thm.\,1]{MR1284231}, but consider all the sequences to be
  depending on the parameter $\lambda\in I$.

  Let $n>N$ and suppose that
  \begin{equation}
    \label{eq:proof-kelley-first}
     v_{n-1}(\lambda)\le X_{n-1}(\lambda)
\le w_{n-1}(\lambda)\,,\qquad \lambda\in I
  \end{equation}
  where $\{X_n(\lambda)\}$ is a solution of (\ref{eq:kelly-eq}), and
  $\{v_n(\lambda)\}$ and $\{w_n(\lambda)\}$ satisfy
  (\ref{eq:kelley-1}) and (\ref{eq:kelley-2}), respectively. Due to
  (\ref{eq:kelley-3}), we have by simple algebraic calculations that, for any
  $\lambda\in I$,
\begin{equation*}
\frac{v_{n-1}(\lambda)}{1+v_{n-1}(\lambda)}\le
\frac{X_{n-1}(\lambda)}{1+X_{n-1}(\lambda)}\le
\frac{w_{n-1}(\lambda)}{1+w_{n-1}(\lambda)}\,.
\end{equation*}
Multiplying these inequalities by $1+\beta_n(\lambda)$, we obtain in
virtue of (\ref{eq:kelley-0}), (\ref{eq:kelley-1}) and
(\ref{eq:kelley-2}), that
\begin{equation*}
   v_{n}(\lambda)\le X_{n}(\lambda)
\le w_{n}(\lambda)\,,\qquad \lambda\in I\,.
\end{equation*}
The proof is completed by induction.
\end{proof}
Note that inside the proof of the previous proposition we justified
that $v_n(\lambda)\le w_n(\lambda)$ for all $n\ge N$.

We shall use Proposition~\ref{thm:kelley1} to obtain the uniform
asymptotics of an increasing solution of the Riccati difference
equation (\ref{eq:kelly-eq}). To obtain a uniform asymptotic expansion
for a decreasing solution of (\ref{eq:kelly-eq}), one has to use the
following result.
\begin{proposition}
\label{thm:kelley2}
Let $I$ be a subset of $\mathbb{R}$.
  Suppose that one can find $N\in\mathbb{N}$ and real sequences
  $\{v_n(\lambda)\}$ and
  $\{w_n(\lambda)\}$ such that
  \begin{align}
   v_n(\lambda)\ge w_n(\lambda)\,,&\qquad\lambda\in I\,,\quad n\ge N\,, \label{eq:cond-kelley2-1}\\
 \sup_{\substack{\lambda\in I\\ n\ge N}}\abs{v_n(\lambda)},&
\sup_{\substack{\lambda\in I\\ n\ge N}}\abs{w_n(\lambda)}<1\,,\label{eq:cond-kelley2-2}
 \end{align}
and (\ref{eq:kelley-0})--(\ref{eq:kelley-2}) hold for $n> N$. 
  Then (\ref{eq:kelly-eq}) has a solution
  $\{X_n(\lambda)\}_{n=N}^\infty$ satisfying
\begin{equation}
\label{eq:satisfying-trap}
    v_n(\lambda)\ge X_n(\lambda)\ge w_n(\lambda)\,,\qquad  n\ge
    N\,,\quad \lambda\in I\,.
  \end{equation}
\end{proposition}
\begin{proof}
 The proof of this assertion follows the proof of
 \cite[Thm.\,2]{MR1284231}, taking into account the dependence on
 $\lambda\in I$.

Let us fix a natural number $s>N$ and define
\begin{equation}
  \label{eq:aux-def-1}
  X_{n,s}(\lambda):=w_n(\lambda)\,,\qquad n\ge s\,.
\end{equation}
By using recurrently the formula
\begin{equation}
  \label{eq:aux-def-2}
  X_{n-1,s}(\lambda):=\frac{X_{n,s}(\lambda)+\beta_n(\lambda)}{1-X_{n,s}(\lambda)}\,,
\end{equation}
for $n=s,s-1,\dots,N+1$, one defines $X_{n,s}(\lambda)$ for all $n\ge
N$. Clearly the sequence $\{X_{n,s}(\lambda)\}_{n=N}^\infty$ satisfies
(\ref{eq:kelly-eq}) for $N< n\le s$.

The inequalities (\ref{eq:kelley-1}) and (\ref{eq:cond-kelley2-2}) imply
\begin{equation}
  \label{eq:bb}
  v_{n-1}(\lambda)\ge\frac{v_n(\lambda)+\beta_n(\lambda)}{1-v_n(\lambda)}\,,
  \qquad n>N\,.
\end{equation}
Analogously, from (\ref{eq:kelley-2}) and (\ref{eq:cond-kelley2-2}),
  one obtains
\begin{equation}
  \label{eq:bbb}
    w_{n-1}(\lambda)\le
    \frac{w_n(\lambda)+\beta_n(\lambda)}{1-w_n(\lambda)}\,,\qquad
    \lambda\in I\,,\quad n>N\,.
  \end{equation}
It follows from (\ref{eq:aux-def-2}), (\ref{eq:bb}), and
(\ref{eq:bbb}) that
\begin{equation}
  \label{eq:trap}
w_n(\lambda)\le X_{n,s}(\lambda)\le v_n(\lambda)\,,\qquad
    \lambda\in I\,,
\end{equation}
for $N\le n<s$. But, due to (\ref{eq:aux-def-1}) and
(\ref{eq:cond-kelley2-1}), the inequality (\ref{eq:trap}) actually
holds for $n\ge N$.

Now, fix a natural number $n\ge N$ and consider the sequence
$\{X_{n,s}(\lambda)\}_{s=s_0}^\infty$, with $s_0>N$. On the basis of
(\ref{eq:bbb}), it can be verified that
$\{X_{n,s}(\lambda)\}_{s=s_0}^\infty$ is a monotone non-decreasing
function and, by (\ref{eq:trap}), it is bounded from above. Then the
sequence has a limit that we denote by $X_n(\lambda)$. This can be
done for any $n\ge N$.  Thus, we have constructed a sequence
$\{X_n(\lambda)\}_{n=N}^\infty$ which satisfies (\ref{eq:kelly-eq})
and is such that (\ref{eq:satisfying-trap}) holds.
\end{proof}
\section{Formal asymptotic analysis of the Riccati equation}
\label{sec:formal-uniform-asymptotics}
In this section we present a heuristic approach for obtaining formal
asymptotic expansions of solutions of the Riccati difference equation
(\ref{eq:kelly-eq}). The asymptotics is constructed term by term
beginning from the leading one. The algorithm given below may be
iterated multiple times until the desired precision. We remark that in
this section there will not be proofs for our asymptotic
formulae. Nevertheless, one may adapt our heuristic to prove
\emph{pointwise}, with respect to $\lambda$, asymptotic
expansions. We are not interested in this pointwise asymptotics since
our approach will yield a stronger result.


Clearly, (\ref{eq:kelly-eq}) can
be written as follows
\begin{equation}
  \label{eq:kelley-multiplied}
  X_n(\lambda)+X_n(\lambda)X_{n-1}(\lambda)=X_{n-1}(\lambda)-\beta_n(\lambda)
\end{equation}
This is the starting point of the algorithm and the first step is to
find the leading term of the asymptotic expansion as $n\to\infty$ of
$\{X_n(\lambda)\}$.  To this end we assume that the sequence
$\{X_n(\lambda)\}$ tend to $0$ as $n\to\infty$ and that it is smooth
with respect to $n$. Then the leading term satisfies the relation
\begin{equation}
  \label{eq:leading-term-Xn}
  X_n(\lambda)+X_n^2(\lambda)=X_n(\lambda)-\beta_n(\lambda)\,.
\end{equation}
Therefore the main term of the asymptotic formula of
$\{X_n(\lambda)\}$ coincides with that of the asymptotic expansion of
$\{\pm\sqrt{-\beta_n(\lambda)}\}$. The next terms of the
expansion may be found by introducing a rectifying sequence
$\{t_n(\lambda)\}$ such that
\begin{equation}
  \label{eq:beta-throu-xt}
  X_n^2(\lambda)=-\beta_n(\lambda) +t_n(\lambda)\,.
\end{equation}
Expressing $\beta_n(\lambda)$ through $X_n(\lambda)$ and
$t_n(\lambda)$ from (\ref{eq:beta-throu-xt}) and substituting it into
(\ref{eq:kelley-multiplied}), one obtains the exact equation
\begin{equation}
  \label{t-n-definition}
  t_n(\lambda)=(X_n(\lambda)-1)(X_n(\lambda)-X_{n-1}(\lambda))\,.
\end{equation}
Since $\beta_n(\lambda)\to 0$ for each $\lambda$, the leading term of
$\{t_n(\lambda)\}$ is given by the leading term of
$\{\pm\left(\sqrt{-\beta_{n}(\lambda)}-\sqrt{-\beta_{n-1}(\lambda)}\right)\}$.

For $\{\beta_n(\lambda)\}$ given by (\ref{eq:beta-x-def}) with
$\{b_n\}$, $\{q_n\}$ defined by (\ref{eq:weights-diagonal}) and
(\ref{complicated-case}), we show in
Appendix~\ref{sec:asymptotics-beta} that the uniform asymptotic
expansion of $\{\beta_n(\lambda)\}$ as $n\to\infty$, when $\lambda>0$, obeys
formula (\ref{eq:beta-asymptotics}).  Thus, by the elementary relation
\begin{equation}
  \label{eq:elementary}
  n^s-(n-1)^s=sn^{s-1} +O(n^{s-2})\quad{as}\quad n\to\infty\,,
\quad s\in\mathbb{R}\,,
\end{equation}
one easily concludes that for each $\lambda>0$
\begin{equation*}
  t_n(\lambda)=\sqrt{\lambda\frac{2^{1-\alpha}}{c_1c_2}}\,
 n^{-1-\alpha/2}
    +o(n^{-1-\alpha/2})\,,\quad\text{ as }\quad n\to\infty\,.
\end{equation*}
We have, therefore, found without proving that for each $\lambda>0$
\begin{equation}
  \label{eq:formal-asymptotics}
  X_n(\lambda)=\pm\sqrt{-\beta_n(\lambda)}+\frac{\alpha}{4n}
+ o(n^{-1})\,,\quad\text{ as }\quad n\to\infty\,.
\end{equation}
To improve the precision of the previous uniform asymptotic formula
one may carry out a second iteration of our heuristic reasoning. To
this end one introduces a second rectifying sequence
$\{u_n(\lambda)\}$ such that
\begin{equation*}
   X_n^2(\lambda)=-\beta_n(\lambda) +\sqrt{\lambda\frac{2^{1-\alpha}}{c_1c_2}}\,
 n^{-1-\alpha/2}+ u_n(\lambda)
\end{equation*}
and repeats what was done before to find an expression for
$t_n(\lambda)$, namely, one expresses $\beta_n(\lambda)$ through
$X_n(\lambda)$ and $u_n(\lambda)$ from the equation above and
substitutes it into (\ref{eq:kelley-multiplied}).

For the purpose of the present work, the precision formally obtained
by the first iteration is sufficient. We remind that our formal
asymptotic expansion will be used only to find the structure of the
majorant and minorant sequences which are fundamental for our uniform
asymptotic method.

\section{Uniform asymptotic analysis of the Riccati equation:
the modified  Kelley method}
\label{sec:kelley estimates}
This section is devoted to the uniform asymptotic analysis of the
solutions of the Riccati equation (\ref{eq:kelly-eq}). To this end, we
construct the majorant and minorant sequences of
Propositions~\ref{thm:kelley1} and \ref{thm:kelley2}.  The formal
asymptotics (\ref{eq:formal-asymptotics}) found in the previous
section and the uniform asymptotic behavior of the sequence
$\{\beta_n(\lambda)\}$, given in (\ref{eq:beta-asymptotics}), will be
at the basis of our considerations.

After reminding the reader on the convention \ref{notation3} of our
notation given in Section~\ref{sec:preliminaries}, let us consider the
following simple and straightforward lemmas.

\begin{lemma}
\label{lem:kelley-maj-min}
Let $\psi_0(\lambda)$ be a uniformly positive and bounded function defined on
$I\subset \mathbb{R}$, i.\,e.,
\begin{equation}
  \label{eq:bounded-separated-from-zero}
  \inf_{\lambda\in I}\psi_0(\lambda)>0\,,\qquad
  \sup_{\lambda\in I}\psi_0(\lambda)<\infty\,.
\end{equation}
 Let $0<p<1$ and $M\in\mathbb{N}$. Suppose that
there is a real sequence $\{\varphi_n(\lambda)\}$ having the following
asymptotic behavior as $n\to\infty$
  \begin{equation}
    \label{eq:master-sequence}
    \varphi_n(\lambda)=
\sum_{k=0}^M\psi_k(\lambda)n^{-s_k}
+\widetilde{o}_I\left(n^{-p-1}\right)\,,\qquad \lambda\in I\,,
  \end{equation}
where $\psi_k(\lambda)$ is bounded for $k=1,\dots,M$ and $p=s_0<s_k\le
p+1$. Then, for any
fixed constants $A^\pm$ obeying
\begin{equation}
\label{eq:parameter-prop1}
   A^+<\frac{p}{2}<A^-\,,
 \end{equation}
there exists $N\in\mathbb{N}$ such that the sequences
 $\{v_n^\pm(\lambda)\}_{n=N}^\infty$ given by
  \begin{equation}
    \label{eq:sequences-prop1}
    v_n^\pm(\lambda):=\pm\varphi_n(\lambda)+
A^\pm n^{-1}\,,\qquad \lambda\in I\,,
\end{equation}
satisfy
\begin{equation}
  \label{eq:ineq-prop1}
  v_n^\pm(\lambda)\left(1+v_{n-1}^\pm(\lambda)\right)
\le v_{n-1}^\pm(\lambda)+\varphi_n^2(\lambda)\,,
\qquad n>N\,,\qquad \lambda\in I\,.
\end{equation}
\end{lemma}
\begin{proof}
  By substituting (\ref{eq:sequences-prop1}) into
  (\ref{eq:ineq-prop1}), one can show that there is a sequence
  $\{\zeta_n(\lambda)\}$ such that the inequality
  (\ref{eq:ineq-prop1}) for $\{v_n^+(\lambda)\}$ is reduced to
\begin{equation*}
   A^+\left(\frac{\varphi_{n-1}(\lambda)}{n} +
      \frac{\varphi_n(\lambda)}{n-1}\right)\le
    \left(\varphi_n(\lambda)-1\right)\left(\varphi_n(\lambda) -
    \varphi_{n-1}(\lambda)\right)+
\zeta_n(\lambda)\,,
\end{equation*}
where $\zeta_n(\lambda)=\widetilde{O}_I(n^{-2})$ as $n\to\infty$. We
allow ourselves to write instead of the previous inequality that
 as $n\to\infty$
  \begin{equation}
    \label{eq:basic-asymp-ineq}
   A^+\left(\frac{\varphi_{n-1}(\lambda)}{n} +
      \frac{\varphi_n(\lambda)}{n-1}\right)\le
    \left(\varphi_n(\lambda)-1\right)\left(\varphi_n(\lambda) -
    \varphi_{n-1}(\lambda)\right)+
\widetilde{O}_I\left(n^{-2}\right)\,,
\end{equation}
hoping that it will not lead to misunderstanding.

Using (\ref{eq:elementary}), one easily obtains the uniform asymptotics
\begin{equation}
  \label{eq:phi-differenciation}
  \varphi_n(\lambda) -
  \varphi_{n-1}(\lambda) = -p\psi_0(\lambda)n^{-p-1} +
\widetilde{o}_I\left(n^{-p-1}\right)\,,\quad{as}\quad n\to\infty\,,
\end{equation}
and then one verifies, after elementary calculations, that
\eqref{eq:basic-asymp-ineq} may be reduced to
\begin{equation}
  \label{eq:asymp-ineq-a}
\psi_0(\lambda)
\left(2A^+-p\right)\le\widetilde{o}_I\left(1\right)\,,
\quad{as}\quad n\to\infty\,,
\end{equation}
for a suitable $\widetilde{o}_I\left(1\right)$ sequence.  In view of
(\ref{eq:asymp-ineq-a}) and the strict positivity of $\psi_0(\lambda)$
for $\lambda\in I$, the first inequality in
(\ref{eq:parameter-prop1}), that is $A^+<\frac{p}{2}$, ensures the
existence of $N_1\in\mathbb{N}$ such that $\{v_n^+(\lambda)\}$
fulfills (\ref{eq:ineq-prop1}) for $n> N_1$. Note that $N_1$ depends
on what constant $A^+$ satisfying the first inequality in
(\ref{eq:parameter-prop1}) has been chosen.

Let us now consider (\ref{eq:ineq-prop1}) for
$\{v_n^-(\lambda)\}$. Calculations analogous to that leading to
(\ref{eq:basic-asymp-ineq}) here yield that (\ref{eq:ineq-prop1}) is
reduced to
 \begin{equation*}
   A^-\left(\frac{\varphi_{n-1}(\lambda)}{n} +
      \frac{\varphi_n(\lambda)}{n-1}\right)\ge
    \left(\varphi_n(\lambda)+1\right)\left(\varphi_{n-1}(\lambda) -
    \varphi_n(\lambda)\right)+
\widetilde{O}_I\left(n^{-2}\right)\,,
\end{equation*}
as $n\to\infty$.  This inequality holds if and only if, for
$\lambda\in I$,
\begin{equation}
  \label{eq:asymp-ineq-a-}
\psi_0(\lambda)
\left(2A^--p\right)\ge\widetilde{o}_I\left(1\right)\,,
\quad{as}\quad n\to\infty\,.
\end{equation}
From (\ref{eq:asymp-ineq-a-}) one concludes that, when the second
inequality in (\ref{eq:parameter-prop1}) holds, that is
$A^->\frac{p}{2}$, there is a number $N_2$ such that
$\{v_n^-(\lambda)\}$ satisfies (\ref{eq:ineq-prop1}) for $n >
N_2$. The number $N_2$ depends on the constant $A^-$ chosen to satisfy the
second inequality in (\ref{eq:parameter-prop1}). The proof is complete
by setting $N=\max\{N_1,N_2\}$.
\end{proof}
\begin{lemma}
  \label{lem:kelley-maj-min2}
  Let the conditions on $\psi_k(\lambda)$ for $k=0,1,\dots,M$ and
  $\{\varphi_n(\lambda)\}$ of Lemma~\ref{lem:kelley-maj-min} be
  satisfied. Then, for any fixed constants $B^\pm$ obeying
\begin{equation}
\label{eq:parameter-prop2}
   B^-<\frac{p}{2}<B^+\,,
 \end{equation}
there exists $N\in\mathbb{N}$ such that the sequences
 $\{w_n^\pm(\lambda)\}_{n=N}^\infty$ given by
  \begin{equation}
    \label{eq:sequences-prop2}
    w_n^\pm(\lambda):=\pm\varphi_n(\lambda)+
B^\pm n^{-1}\,,\qquad \lambda\in I\,,
\end{equation}
satisfy
\begin{equation}
  \label{eq:ineq-prop2}
  w_n^\pm(\lambda)\left(1+w_{n-1}^\pm(\lambda)\right)
\ge w_{n-1}^\pm(\lambda)+\varphi_n^2(\lambda)\,,
\qquad n>N\,,\qquad \lambda\in I\,.
\end{equation}
\end{lemma}
\begin{proof}
  The proof repeats the reasoning of that of Lemma~\ref{lem:kelley-maj-min}.
\end{proof}
We draw the reader's attention to the following. First, in
(\ref{eq:sequences-prop1}) and (\ref{eq:sequences-prop2}) we have
reproduced the structure of the formal asymptotics of solutions of the
Riccati equation (\ref{eq:formal-asymptotics}). Second, if the
asymptotic expansion of $\{\varphi_n(\lambda)\}$ had been given in a
less precise form than (\ref{eq:master-sequence}), it would not have
been sufficient for proving that (\ref{eq:phi-differenciation}) holds
true and, then, for proving the assertions of
Lemmas~\ref{lem:kelley-maj-min} and \ref{lem:kelley-maj-min2}. Note
that, for obtaining a condition on $A^+$, the order of the
leading term in the right-hand side of (\ref{eq:basic-asymp-ineq})
must coincide with that of the left-hand side. The expansion
(\ref{eq:phi-differenciation}) ensures this, but for
(\ref{eq:phi-differenciation}) to hold, one needs to know that all
terms of the asymptotic expansion of $\{\varphi_n(\lambda)\}$ are
``differentiable with respect to $n$'' up to the precision indicated by
(\ref{eq:master-sequence}). Clearly, it is not important to know the
concrete form of the functions $\psi_k(\lambda)$, $k=0,1,\dots,M$ as
long as they satisfy the conditions of Lemmas~\ref{lem:kelley-maj-min}
and \ref{lem:kelley-maj-min2}.

Bellow we shall show that the majorant and minorant sequences referred
at the beginning of this section are constructed from the sequences
$\{v_n^\pm(\lambda)\}$, $\{w_n^\pm(\lambda)\}$ under the assumption
that
\begin{equation}
  \label{eq:phi-function-def}
  \varphi_n(\lambda)=\sqrt{-\beta_n(\lambda)}\,.
\end{equation}
Thus, (\ref{eq:master-sequence}) shows how precise our calculations of
the asymptotic expansion of $\{\beta_n(\lambda)\}$, and consequently
of the Poincar\'{e} coefficients, should be.  Note that, by
considering the remark at the end of the previous paragraph, we are
not concerned with the concrete form of the corresponding functions
$\psi_k(\lambda)$, $k=1,\dots,M$, stemming from
(\ref{eq:phi-function-def}).

\begin{lemma}
 \label{lem:majorant-minorant-proof}
 Let (\ref{eq:alpha-def}) and (\ref{eq:c1-c2-positivity}) hold and let
 $\{\beta_n(\lambda)\}$ have the asymptotic behavior given by
 (\ref{eq:beta-asymptotics}). Suppose that $A^\pm$ and $B^\pm$ obey
 (\ref{eq:parameter-prop1}) and (\ref{eq:parameter-prop2}). Then, for
 any bounded subset $I\subset\mathbb{R}_+$ separated from zero, the
 sequences $\{v_n^+(\lambda)\}$, $\{w_n^+(\lambda)\}$ and
 $\{v_n^-(\lambda)\}$, $\{w_n^-(\lambda)\}$, given by
 (\ref{eq:sequences-prop1}) and (\ref{eq:sequences-prop2}) with
 (\ref{eq:phi-function-def}), satisfy the conditions
 of Propositions \ref{thm:kelley1} and \ref{thm:kelley2},
 respectively.
\end{lemma}
\begin{proof}
First note that (\ref{eq:beta-asymptotics}) implies the existence of
$n_1\in\mathbb{N}$ such that
\begin{equation}
  \label{eq:positivity-inequalities}
  \inf_{\lambda\in I}\{ 1+v_n^\pm(\lambda)\}>0\,,
\qquad \inf_{\lambda\in I}\{1+w_n^\pm(\lambda)\}>0\,,\qquad n\ge n_1\,.
\end{equation}
Thus, \eqref{eq:kelley-3} holds for $n\ge n_1$.

Now, in view of the asymptotic expansion (\ref{eq:beta-asymptotics}),
the sequence $\{\varphi_n(\lambda)\}$ satisfies the conditions
required by Lemmas~\ref{lem:kelley-maj-min} and
\ref{lem:kelley-maj-min2}. Indeed, on the one hand, the condition on
$p$ holds due to (\ref{eq:alpha-def}) since $p=\frac{\alpha}{2}$. On
the other hand, a simple computation shows that
\begin{equation*}
  \psi_0(\lambda)=\sqrt{\lambda\frac{2^{1-\alpha}}{c_1c_2}}\,,\qquad
\lambda\in I\,.
\end{equation*}
Thus, taking into account that the bounded set $I\subset\mathbb{R}_+$
is separated from zero, (\ref{eq:c1-c2-positivity}) implies that
$\psi_0(\lambda)$ satisfies (\ref{eq:bounded-separated-from-zero}). The
conditions on $\psi_k(\lambda)$, $k=1,\dots,M$, are easily verified.

By Lemmas~\ref{lem:kelley-maj-min} and \ref{lem:kelley-maj-min2} there
is a natural number $n_2\ge n_1$ such that
\begin{align}
  v_n^\pm(\lambda)\left(1+v_{n-1}^\pm(\lambda)\right)&
\le v_{n-1}^\pm(\lambda)-\beta_n(\lambda)\,,
\qquad n>n_2\,,\qquad \lambda\in I\,,\label{eq:minorant}\\
  w_n^\pm(\lambda)\left(1+w_{n-1}^\pm(\lambda)\right)&
\ge w_{n-1}^\pm(\lambda)-\beta_n(\lambda)\,,
\qquad
n>n_2\,,\qquad \lambda\in I\,.\label{eq:majorant}
\end{align}
Due to (\ref{eq:positivity-inequalities}), the inequalities
(\ref{eq:minorant}) and (\ref{eq:majorant}) are, respectively,
equivalent to (\ref{eq:kelley-1}) and (\ref{eq:kelley-2}) for $n\ge
n_2$. On the other hand, (\ref{eq:beta-asymptotics}) implies the
fulfilment of (\ref{eq:kelley-0}) for some $n_3$ independent of
$\lambda$. We choose $n_3\ge n_2$. After observing that
$v_{n_3}^+(\lambda)<w_{n_3}^+(\lambda)$ for all $\lambda\in I$, one
can conclude that $\{v_n^+(\lambda)\}$ and $\{w_n^+(\lambda)\}$
satisfy the conditions of Proposition~\ref{thm:kelley1} with $N=n_3$.

On the basis of \eqref{eq:beta-asymptotics}, one easily concludes
that there is $n_4\in\mathbb{N}$ such that $\{v_n^-(\lambda)\}$ and
$\{w_n^-(\lambda)\}$ satisfy \eqref{eq:cond-kelley2-1} and
\eqref{eq:cond-kelley2-2}, respectively, for $n\ge n_4$. Since
(\ref{eq:kelley-4})--(\ref{eq:kelley-1}) are also fulfilled for some
$n_5\ge n_4$, one
verifies that $\{v_n^-(\lambda)\}$ and $\{w_n^-(\lambda)\}$ obey the
conditions of Proposition~\ref{thm:kelley2} with $N=n_5$.
\end{proof}
The following result gives the uniform asymptotics of solutions
of the Riccati equation (\ref{eq:kelly-eq}).
\begin{theorem}
  \label{prop:x-asymptotics}
  Let (\ref{eq:alpha-def}) and (\ref{eq:c1-c2-positivity}) hold, and
  suppose that $\{\beta_n(\lambda)\}$ has the asymptotic behavior
  given by (\ref{eq:beta-asymptotics}). Then, for any bounded set
  $I\subset\mathbb{R}_+$ separated from zero, there is
  $N\in\mathbb{N}$ such that there are
  solutions $\{X_n^\pm(\lambda)\}_{n=N}^\infty$ of (\ref{eq:kelly-eq})
  for $n\ge N$ having the following asymptotic behavior
  \begin{equation}
    \label{eq:x-solution}
    X_n^\pm(\lambda)=\pm\sqrt{-\beta_n(\lambda)}+
    \widetilde{O}_I\left(n^{-1}\right)\,\quad\text{ as }\, n\to\infty\,.
  \end{equation}
\end{theorem}
\begin{proof}
  Consider the sequences $\{v_n^\pm(\lambda)\}$ and
  $\{w_n^\pm(\lambda)\}$ given by
  (\ref{eq:sequences-prop1}) and (\ref{eq:sequences-prop2}). By
  Lemma~\ref{lem:majorant-minorant-proof} and
  Proposition~\ref{thm:kelley1} there is a constant $N_1\in\mathbb{N}$
  and a solution $\{X_n^+(\lambda)\}$ of (\ref{eq:kelly-eq}) such
  that, for
  all $\lambda\in I$,
  \begin{equation*}
    v_n^+(\lambda)\le X_n^+(\lambda)\le w_n^+(\lambda)\,,\qquad n\ge N_1\,.
  \end{equation*}
Due to (\ref{eq:sequences-prop1}) and (\ref{eq:sequences-prop2})
\begin{equation*}
  w_n^+(\lambda)-v_n^+(\lambda)=\left(B^+-A^+\right)n^{-1}\,.
\end{equation*}
Similarly, by Lemma~\ref{lem:majorant-minorant-proof} and
Proposition~\ref{thm:kelley2}, one can find $N_2\in\mathbb{N}$ and a
solution $\{X_n^-(\lambda)\}$ of (\ref{eq:kelly-eq}) such that for all
$\lambda\in I$
\begin{equation*}
      v_n^-(\lambda)\ge X_2^-(\lambda)\ge w_n^-(\lambda)\,,
\qquad n\ge N_2\,.
\end{equation*}
The proof is complete by noting that
\begin{equation*}
  v_n^-(\lambda)-w_n^-(\lambda)=\left(A^--B^-\right)n^{-1}
\end{equation*}
and setting $N=\max\{N_1,N_2\}$.
\end{proof}
\section{Uniform asymptotics of the generalized
  eigenvectors}
\label{sec:generalized-eigenvectors}
The results of the previous section allow us to obtain the uniform
asymptotic behavior of the generalized eigenvectors corresponding to
the family of Jacobi matrices $J(c_1,c_2)$ introduced in
Section~\ref{sec:preliminaries}. Our results are restricted to the
case given by (\ref{eq:c1-c2-positivity}) and (\ref{complicated-case})
and illustrated in Figure~\ref{fig.2}.

We begin by finding the asymptotic behavior of the solutions of the
Poincar\'{e} type equation (\ref{eq:birkhoff-adams-eq-x}).

\begin{theorem}
  \label{thm:x-asymptotics}
  Let $\{q_n\}_{n=1}^\infty$ and
$\{b_n\}_{n=1}^\infty$ be defined by (\ref{eq:weights-diagonal}) with
(\ref{eq:alpha-def}), (\ref{eq:c1-c2-positivity}) and
(\ref{complicated-case}). Also let $I$ be any  bounded subset of
$\mathbb{R}_+$ separated from zero. Then, there is $N\in\mathbb{N}$
such that there are solutions $\{x_n^\pm(\lambda)\}_{n=N}^\infty$ of
(\ref{eq:birkhoff-adams-eq-x}) for $n> N$, with $\{F_n(\lambda)\}$ and
$\{G_n(\lambda)\}$ given by (\ref{eq:f-g-definition}), having
  the following uniform asymptotic behavior as $n\to\infty$
  \begin{equation}
    \label{eq:xi-solutions}
    x_n^\pm(\lambda)=\exp\left(\pm\frac{\sqrt{\lambda\frac{2^{1-\alpha}}{c_1c_2}}}
    {1-\frac{\alpha}{2}}n^{1-\frac{\alpha}{2}}+
    \widetilde{O}_{I}(n^{1-\alpha})\right)\,.
  \end{equation}
\end{theorem}
\begin{proof}
   Since
  (\ref{eq:alpha-def}) and (\ref{complicated-case}) hold, the
  asymptotic expansion of $\{F_n(\lambda)\}$ is given by
  (\ref{eq:F-n-expansion}). This implies that there is $N_1\in\mathbb{N}$ such
  that the conditions of Lemma \ref{lem:Kelley-equation} are
  satisfied (see Remark~\ref{rem:riccati}).
 Then the solutions of (\ref{eq:birkhoff-adams-eq-x}) are
  given by the solutions of (\ref{eq:kelly-eq}) for $n>N_1$ and
  $x_n(\lambda)\ne 0$. According to (\ref{eq:x-definition}), the sequence
  $\{x_n(\lambda)\}_{n=N_1}^\infty$ satisfies
 \begin{equation*}
   \frac{x_{n+1}(\lambda)}{x_{n}(\lambda)}=
   (-1)\frac{F_n(\lambda)}{2}\left(1+X_n(\lambda)\right)\,,
   \qquad n\ge N_1,
 \end{equation*}
where $\{X_n(\lambda)\}_{n=N_1}^\infty$ is a solution of
(\ref{eq:kelly-eq}) and $x_n(\lambda)\ne 0$ for $n>N_1$.
Hence,
\begin{equation}
\label{eq:emergency}
  x_n(\lambda)=x_{N_1}(\lambda)(-1)^{n-N_1}
\prod_{k=N_1}^{n-1}\left[
\frac{F_k(\lambda)}{2}\left(1+X_k(\lambda)\right)\right]\,,
\end{equation}
with $x_{N_1}(\lambda)\ne 0$ for $\lambda\in I$.
Let us find the asymptotic behavior of the product.
We assume that $N_2\ge N_1$ is such that
\begin{equation*}
  \sup_{\lambda\in I}\abs{F_n(\lambda)-2}<\frac{1}{2}\,,\qquad
\sup_{\lambda\in I}\abs{X_n(\lambda)}<\frac{1}{2}\quad n\ge N_2
\end{equation*}
This can be done by virtue of (\ref{eq:F-n-expansion}) and
(\ref{eq:x-solution}) taking into account
(\ref{eq:beta-asymptotics}).

 We have
\begin{equation*}
\prod_{k=N_2}^{n-1} \frac{F_k(\lambda)}{2}=\exp
\left(\sum_{k=N_2}^{n-1}
  \log\left(\frac{F_k(\lambda)}{2}\right)\right)\,.
\end{equation*}
On the basis of (\ref{eq:F-n-expansion}), one verifies that
\begin{equation}
  \label{eq:product-F}
\prod_{k=N_2}^{n-1} \frac{F_k(\lambda)}{2}
=\exp\widetilde{O}_I(n^{1-\alpha})\,,\qquad\text{as }\quad n\to\infty\,.
\end{equation}
On the other hand, by (\ref{eq:alpha-def}),
(\ref{eq:c1-c2-positivity}) and (\ref{complicated-case}), the
asymptotic expansion (\ref{eq:beta-asymptotics}) holds and the
conditions of Theorem~\ref{prop:x-asymptotics} are met. Thus, there is
$N_3\ge N_2$ such that (\ref{eq:kelly-eq}) has solutions
(\ref{eq:x-solution}) for $n>N_3$. Now,
 in view of \cite[Sec.\,2]{MR1284231},
\begin{equation}
\label{eq:product-X}
  \prod_{k=N_3}^{n-1}\left( 1+X_k^\pm(\lambda)\right)
=K_n(\lambda)\exp\left(\sum_{k=N_3}^{n-1}
\sum_{j=1}^s\frac{(-1)^{j-1}}{j}\left(X_k^\pm(\lambda)\right)^j\right)\,,
\end{equation}
where $K_n(\lambda)\to K(\lambda)>K>0$, as $n\to\infty$, and $s$ has
been chosen so that $\sup_{\lambda\in I}\sum_{k=N_3}^\infty
\left(X_k^\pm(\lambda)\right)^{s+1}<\infty$. We can always find such
$s$ in view of (\ref{eq:x-solution}) and
(\ref{eq:beta-asymptotics}). Straightforward computations yield the
following asymptotic behavior
\begin{equation*}
  \prod_{k=N_3}^{n-1}\left( 1+X^\pm_k(\lambda)\right)=
  \exp\left(\pm\frac{\sqrt{\lambda\frac{2^{1-\alpha}}{c_1c_2}}}
    {1-\frac{\alpha}{2}}n^{1-\frac{\alpha}{2}}+
    \widetilde{O}_I(n^{1-\frac{3}{2}\alpha}) \right)\,,\qquad\text{as }\quad n\to\infty\,.
\end{equation*}
For the proof to be complete, set $N:=N_3$.
\end{proof}
\begin{remark}
  We could have obtained a more precise asymptotic formula than
  (\ref{eq:xi-solutions}). This may be done by using an asymptotic
  expansion of $\{\beta_n(\lambda)\}$ more precise than
  (\ref{eq:beta-asymptotics}) and refining the results of
  Section~\ref{sec:kelley estimates}. However, for our goal, the
  precision of the asymptotics (\ref{eq:xi-solutions}) is sufficient.
\end{remark}
\begin{corollary}
  \label{prop:solutions-asympt}
  Let $\{q_n\}_{n=1}^\infty$ and $\{b_n\}_{n=1}^\infty$ be defined by
  (\ref{eq:weights-diagonal}) with (\ref{eq:alpha-def}),
  (\ref{eq:c1-c2-positivity}) and (\ref{complicated-case}). Suppose
  that $I$ is a bounded subset of $\mathbb{R}_+$ separated from
  zero. Then, there is $N\in\mathbb{N}$ such that there are linearly
  independent solutions $\{f_n^\pm(\lambda)\}_{n=N}^\infty$ of
  (\ref{eq:main-recurrence}) for $n> N$ having the following uniform
  asymptotic behavior as $n\to\infty$
  \begin{equation}
    \label{eq:gen-eigen-solutions}
     f_n^\pm(\lambda)=
    \exp\left(\pm \frac{\sqrt{\frac{\lambda}{2c_1c_2}}}
    {1-\frac{\alpha}{2}}n^{1-\frac{\alpha}{2}}+
\widetilde{O}_{I}(n^{1-\alpha})\right).
  \end{equation}
\end{corollary}
\begin{proof}
  By Remark~\ref{rem:rem}, the solution
  $\{x_n^+(\lambda)\}$, respectively $\{x_n^-(\lambda)\}$, of
  (\ref{eq:birkhoff-adams-eq-x}) found in
  Theorem~\ref{thm:x-asymptotics} determines uniquely a solution
$\{y_n^+(\lambda)\}$, respectively $\{y_n^-(\lambda)\}$, of
  (\ref{eq:birkhoff-adams-eq-y}). By
  (\ref{eq:all_determined_first_eq}) we have the following asymptotic formula
  as $n\to\infty$
\begin{equation*}
  y_n^\pm(\lambda)=
    \exp\left(\pm\frac{\sqrt{\lambda\frac{2^{1-\alpha}}{c_1c_2}}}
    {1-\frac{\alpha}{2}}n^{1-\frac{\alpha}{2}}+
\widetilde{O}_{I}(n^{1-\alpha})\right).
\end{equation*}
The solutions of (\ref{eq:main-recurrence}) for $n>N$ are obtained
from the sequences $\{x_n^\pm(\lambda)\}$ and $\{y_n^\pm(\lambda)\}$
by means of (\ref{eq:x-y-sequences}). Clearly, $\{f^+_n(\lambda)\}$
and $\{f^-_n(\lambda)\}$ are linearly independent and every solution
of (\ref{eq:main-recurrence}), for $n>N$, is a linear combination of
these sequences.
\end{proof}
\begin{remark}
\label{rem:last}
  The solutions $\{f_n^\pm(\lambda)\}_{n=N}^\infty$ of
  (\ref{eq:main-recurrence}) for $n> N$ given by
  Corollary~\ref{prop:solutions-asympt} may be extended by
  Remark~\ref{rem:rem} to sequences
  $\{f_n^\pm(\lambda)\}_{n=1}^\infty$ being linearly independent
  solutions of (\ref{eq:main-recurrence}) for $n>1$. All solutions of
  (\ref{eq:main-recurrence}) for $n>1$ are linear combinations of
  $\{f_n^\pm(\lambda)\}_{n=1}^\infty$, in particular, the generalized
  eigenvectors of $J$ (see the Introduction).
\end{remark}
\begin{corollary}
  \label{cor:last-cor}
  Let $\{q_n\}_{n=1}^\infty$ and $\{b_n\}_{n=1}^\infty$ be defined by
  (\ref{eq:weights-diagonal}) with (\ref{eq:alpha-def}),
  (\ref{eq:c1-c2-positivity}) and (\ref{complicated-case}). Suppose
  that $I$ is a bounded subset of $\mathbb{R}_+$ separated from
  zero and that $\lambda_0\in\sigma_{pp}(J)\cap I$. Then the
  sequence
  \begin{equation*}
 \{f_n^-(\lambda_0)\}_{n=1}^\infty\,,
  \end{equation*}
  where $\{f_n^-(\lambda)\}_{n=1}^\infty$ is the sequence given in
  Remark \ref{rem:last}, is the eigenvector of $J$ corresponding to
  $\lambda_0$. Moreover, there is a uniformly separated from zero and
  bounded function
  \begin{equation*}
    \mathcal{C}:\sigma_{pp}(J)\cap I\to\mathbb{R}
  \end{equation*}
such that
\begin{equation}
  \label{eq:poly-first-kind}
  f_n^*(\lambda)=\mathcal{C}(\lambda)f_n^-(\lambda)\,,\quad
\lambda\in\sigma_{pp}(J)\cap I\,,
\end{equation}
where $\{f_n^*(\lambda)\}$ is the sequence of orthogonal polynomials of
the first kind associated to $J$ (see the Introduction).
\end{corollary}
\begin{proof}
  Consider equation (\ref{eq:main-recurrence}).  Assign
  $f_1^*(\lambda):=1$ for all $\lambda$ and
  $f_2^*(\lambda):=\frac{\lambda-q_1}{b_1}$ By using recurrently
  (\ref{eq:main-recurrence}), one defines the sequence
  $\{f_n^*(\lambda)\}_{n=1}^\infty$ of orthogonal polynomials of the
  first kind associated with $J$. Clearly, evaluation of this sequence in
  $\lambda_0$ yields an eigenvector.

  Now, since $\lambda_0\in\sigma_{pp}(\lambda)$ the solution
  $\{f_n^*(\lambda_0)\}_{n=1}^\infty$ of (\ref{eq:main-recurrence}) is
  decaying and therefore it is equal to
  $\{f_n^-(\lambda_0)\}_{n=1}^\infty$ modulo a constant factor. From
  this, one obtains the first assertion of the corollary and
  (\ref{eq:poly-first-kind}). The stated properties of the function
  $\mathcal{C}$ follow from the boundedness of $f_n^*(\lambda)$ with
  respect to $\lambda\in I$ for and fixed $n$, the asymptotic formula
  (\ref{eq:gen-eigen-solutions}), and the fact that $\mathcal{C}$ is
  independent of $n$. Indeed, fix a natural number $n$ sufficiently
  large so that \eqref{eq:gen-eigen-solutions} holds, then
  $f_{n}^-(\lambda)$ is uniformly positive and bounded for $\lambda\in
  I$. Thus, by the boundedness of the polynomials of the first kind,
  we obtain that $\mathcal{C}$ is a uniformly separated from zero and
  bounded function for $\lambda$ belonging to $\sigma_{pp}(J)\cap I$.

  Note that under our considerations the roots of the polynomial
  $f_n^*(\lambda)$ for any sufficiently large $n\in\mathbb{N}$ are not
  in $\sigma_{pp}(J)$.
\end{proof}
\section{Discrete spectrum}
\label{sec:discrete-spectrum}
The main results of the previous section indicate, among other things,
that the sequence $\{f_n^-(\lambda)\}_{n=1}^\infty$, given in
Remark~\ref{rem:last}, is a subordinate solution
\cite{MR915965,MR1179528} of the recurrence equation
(\ref{eq:main-recurrence}) for $\lambda\in I$, where
$\{q_n\}_{n=1}^\infty$ and $\{b_n\}_{n=1}^\infty$ are given by
(\ref{eq:weights-diagonal}), (\ref{eq:alpha-def}),
(\ref{eq:c1-c2-positivity}), (\ref{complicated-case}), and $I$ is a
bounded subset of $\mathbb{R}_+$ separated from zero.  Thus, by
invoking \cite{MR915965,MR1179528}, we arrive at the following
assertion.
\begin{theorem}
  \label{prop:pure-point}
  Let the sequences $\{q_n\}_{n=1}^\infty$ and $\{b_n\}_{n=1}^\infty$
  be given by (\ref{eq:weights-diagonal}) with (\ref{eq:alpha-def}),
  (\ref{eq:c1-c2-positivity}), and (\ref{complicated-case}). Then the
  spectrum of $J$ in $\mathbb{R}_+$ is pure point.
\end{theorem}

We expect the spectrum to be discrete in $\mathbb{R}$, that is finite
in $I$, on the grounds that the uniform estimate of the asymptotic
remainder allows, ``in some sense'', to avoid dealing with the
generalized eigenvector's tail. This very informal reasoning provides
a clue to the problem of estimates for the number of eigenvalues,
namely by recurring to \cite[Sec. 3 Thm. 6]{MR0185471}.  Below we show
that this speculation on the discreteness of the spectrum is indeed
true: we prove absence of accumulation points of $\sigma_{pp}(J)$ in
any closed bounded interval of $\mathbb{R}_+$. To this end we make use
of a technique developed in \cite{silva-otamp} which, in its turn,
relies on ideas put forth in \cite{MR1137522,MR0178362}. The main
ingredients of this recipe are the uniform asymptotic formulae found
in the previous section and Corollary \ref{cor:last-cor}.

We shall need the following auxiliary result.
\begin{lemma}
  \label{lem:uniform-convergence}
For any closed bounded interval $I$ of $\mathbb{R}_+$and any $\epsilon>0$,
there exists $K\in\mathbb{N}$ such that
\begin{equation*}
   \sup_{\lambda\in I}\sum_{n=K}^\infty
    \abs{f_n^-(\lambda)}^2<\epsilon\,,
\end{equation*}
where $\{f^-_n(\lambda)\}$ is the sequence given in Remark~\ref{rem:last}.
\end{lemma}
\begin{proof}
  The assertion follows straightforwardly from the uniform asymptotics
  (\ref{eq:gen-eigen-solutions}).
\end{proof}
\begin{theorem}
  \label{prop:discrete-spectrum}
  Let the sequences $\{q_n\}$ and $\{b_n\}$ be given by
  (\ref{eq:weights-diagonal}) with (\ref{eq:alpha-def}),
  (\ref{eq:c1-c2-positivity}) and (\ref{complicated-case}). Then the
  spectrum of $J$ in $\mathbb{R}_+$ is discrete, i.\,e.,
  \begin{equation*}
    \sigma(J)\cap \mathbb{R}_+=\sigma_{disc}(J)\cap \mathbb{R}_+\,.
  \end{equation*}
\end{theorem}
\begin{proof}
  By Theorem~\ref{prop:pure-point} we know that the spectrum is pure
  point in any closed bounded interval $I$ of $\mathbb{R}_+$.  Suppose
  that $\sigma(J)$ has a point of accumulation $\mu$ in the interior
  of $I$. Let $\lambda$ and $\lambda'$ ($\lambda\ne\lambda'$) be
  arbitrarily chosen from $\sigma(J)\cap V_{\delta}(\mu)$, where
  $V_{\delta}(\mu)$ is a $\delta$-neighborhood of $\mu$, and $\delta$
  is so small that $V_{\delta}(\mu)\subset I$. Of course, $\mu$ need
  not be itself an eigenvalue.

For any $K\in\mathbb{N}$
  the following inequality holds
  \begin{equation}
 \label{eq:orthogonal-ineq}
    \begin{split}
      \abs{(f^*(\lambda),f^*(\lambda'))_{l_2(\mathbb{N})}}&= \abs{
        \sum_{n=1}^\infty
        f^*_n(\lambda)\overline{f^*_n(\lambda')}}\geq\\&\geq \abs{
        \sum_{n=1}^{K}f^*_n(\lambda)\overline{f^*_n(\lambda')}} -
      \abs{ \sum_{n>K} f^*_n(\lambda)\overline{f^*_n(\lambda')}}\,,
 \end{split}
\end{equation}
where $\{f_n^*(\lambda)\}$ is the sequence of polynomials of the first
kind associated with $J$.

Let us estimate the last term in the right hand side of
(\ref{eq:orthogonal-ineq}). To this end express $f_n^*(\lambda)$ and
$f_n^*(\lambda')$ through (\ref{eq:poly-first-kind}). Now, on the
basis of Corollary \ref{cor:last-cor}, namely the boundedness of
$\mathcal{C}$, we verify that Lemma~\ref{lem:uniform-convergence} and
the Cauchy-Schwartz inequality allow us to choose $K$ independent of
$\lambda,\lambda'\in I$, so that
\begin{equation}
\label{eq:last-term}
   \abs{ \sum_{n>K} f^*_n(\lambda)\overline{f^*_n(\lambda')}} <
   \frac{1}{2}\,.
\end{equation}
Now, consider the first term in the right-hand side of
(\ref{eq:orthogonal-ineq}). We have
\begin{equation}
  \label{eq:ineq-one}
 \begin{split}
   \abs{\sum_{n=1}^{K}f^*_n(\lambda)\overline{f^*_n(\lambda')}}
   &\geq \sum_{n=1}^{K}\abs{f^*_n(\lambda)}^2 -
   \abs{\sum_{n=1}^{K}f^*_n(\lambda)
     (\overline{f^*_n(\lambda')-f^*_n(\lambda)})}
   \\ &\geq 1 -
   \abs{\sum_{n=1}^{K}f^*_n(\lambda)
     (\overline{f^*_n(\lambda')-f^*_n(\lambda)})}\,,
 \end{split}
\end{equation}
due to the fact that $f_1^*(\lambda)\equiv 1$. Using the inequality
$\abs{\lambda'-\lambda}<2\delta$, which holds since
$\lambda,\lambda'\in V_\delta(\mu)$, one may write
\begin{equation*}
  \abs{\sum_{n=1}^{K}f^*_n(\lambda)
    (\overline{f^*_n(\lambda')-f^*_n(\lambda)})}
  \leq \max_{1\leq n\leq K}
  \omega_n(2\delta)\sum_{n=1}^{K}\abs{f^*_n(\lambda)}\,,
\end{equation*}
where
\begin{equation*}
  \omega_n(2\delta)=
\sup_{\substack{\abs{\lambda'-\lambda} < 2\delta \\ \lambda',
    \,\lambda\in I}}\abs{f^*_n(\lambda')-f^*_n(\lambda)}
\end{equation*}
is the modulus of continuity of $f_n^*(\lambda)$ on $I$. Thus, by the
uniform continuity of $\{f_n^*(\lambda)\}$ and its uniform
boundedness, which follows from the uniform continuity and the
boundedness for a fixed $\lambda$, one may take $\delta$ sufficiently
small so that
\begin{equation}
  \label{eq:middle-term}
  \abs{\sum_{n=1}^{K}f^*_n(\lambda)
    (\overline{f_n^*(\lambda')-f^*_n(\lambda)})}
  < \frac{1}{2}\,.
\end{equation}
From (\ref{eq:orthogonal-ineq}), (\ref{eq:last-term}), (\ref{eq:ineq-one})
and (\ref{eq:middle-term}), we conclude that
\begin{equation*}
  (f^*(\lambda),f^*(\lambda'))_{l^2(\mathbb{N})} >
  1 -  \frac{1}{2}-\frac{1}{2} = 0\,.
\end{equation*}
If $\lambda$ and $\lambda'$ are in $\sigma_{pp}(J)$, it must be that
$f^*(\lambda)\perp f^*(\lambda')$ since $J=J^*$. This is in
contradiction with the above inequality and thus with the accumulation
of eigenvalues of $J$ to $\mu$.
\end{proof}
We conclude this section with the following comment. The proof of the
spectrum's discreteness crucially relies not only on the uniform
asymptotics of $\{f_n^-(\lambda)\}$, but also on the uniform
continuity of the elements of $\{f_n^*(\lambda)\}$. Corollary
\ref{cor:last-cor} allows to properly ``connect'' this two solutions
and take advantage of the properties of each sequence. Note that this
``proper connection'' is possible due to the fact that the asymptotic
expansion of the decaying solution is uniformly separated from zero
for $\lambda\in I\cap\sigma_{pp}(J)$ and all fixed sufficiently large
$n\in\mathbb{N}$.


\appendix
\begin{center}
  \textbf{\Large Appendix}
\end{center}
\section{Asymptotics of the Poincar\'{e} coefficients}
\label{sec:asymptotics-coef}
Let us study the asymptotic behavior of the sequences
$\{F_n(\lambda)\}$ and $\{G_n(\lambda)\}$, given by (\ref{eq:f-g-definition}), as
$n\to\infty$.  We assume that the sequences $\{q_n\}_{n=1}^\infty$ and
$\{b_n\}_{n=1}^\infty$ given by (\ref{eq:weights-diagonal}) with
(\ref{eq:alpha-def}) and (\ref{complicated-case}). Let $I$ be
any bounded interval of $\mathbb{R}_+$. Take a natural number $N\ge
\frac{(\sup I)^{1/\alpha}}{2}$. As the definition of $F_n(\lambda)$
requires, we have $q_{2n}\not\in I$ as soon as $n\ge N$.

Let us substitute the expression for $\{q_n\}_{n=1}^\infty$ and
$\{b_n\}_{n=1}^\infty$ given by (\ref{eq:weights-diagonal})
into (\ref{eq:f-g-definition}). We then write
\begin{equation*}
  F_n(\lambda)=\frac{c_2}{c_1}A_n(\lambda)
  -\frac{1}{c_1c_2}B_n(\lambda)+\frac{c_1}{c_2}C_n\,,
\qquad \lambda\in I\,,\quad n\ge N\,,
\end{equation*}
where, for any $\lambda\in I$, $n\ge N$
\begin{align*}
A_n(\lambda)&:=\frac{((2n+2)^\alpha-\lambda)(2n)^{2\alpha}}
{((2n)^\alpha-\lambda)(2n+1)^\alpha(2n+2)^\alpha}\\
B_n(\lambda)&:=\frac{((2n+1)^\alpha-\lambda)((2n+2)^\alpha-\lambda)}
{(2n+1)^\alpha(2n+2)^\alpha}\\
C_n&:=\frac{(2n+1)^\alpha}{(2n+2)^\alpha}
\end{align*}
First we obtain asymptotic formulae for $\{A_n(\lambda)\}$,
$\{B_n(\lambda)\}$, and $\{C_n\}$ separately.

Since $\lambda$ is confined in the finite interval $I$, we have the
following \emph{uniform} asymptotic expansion
\begin{equation*}
  \frac{1}{(2n)^\alpha-\lambda}=\frac{1}{(2n)^\alpha}
\left(1+\frac{\lambda}{(2n)^\alpha} +
\frac{\lambda^2}{(2n)^{2\alpha}} +
\widetilde{O}_I(n^{-3\alpha})\right)\,,
\quad\text{ as}\quad n\to\infty\,.
\end{equation*}
Using the fact that
$\frac{(2n)^{2\alpha}}{(2n+1)^\alpha(2n+2)^\alpha}=1-\frac{3\alpha}{2n}
+ O(n^{-2})$ as $n\to\infty$, one obtains
{\small
\begin{equation*}
  A_n(\lambda)=\frac{(2n+2)^\alpha\!-\!\lambda}{(2n)^\alpha}
\left(1+\frac{\lambda}{(2n)^\alpha} +
\frac{\lambda^2}{(2n)^{2\alpha}} +
\widetilde{O}_I(n^{-3\alpha})\right)\!\!
\left(1 -\frac{3\alpha}{2n}+ \widetilde{O}_I(n^{-2})\right)
\end{equation*}
}
Hence, taking into account (\ref{eq:alpha-def}), one has
\begin{equation*}
   A_n(\lambda)=1-\frac{\alpha}{2n}+\widetilde{O}_I(n^{-1-\alpha})\,,
\quad\text{ as}\quad n\to\infty\,.
\end{equation*}
By observing that $B_n(\lambda)=A_n(\lambda)\left(\frac{(2n+1)^\alpha
    -\lambda}{(2n)^\alpha}\right)\left(1-\frac{\lambda}{(2n)^\alpha}\right)$,
one easily verifies that
\begin{equation*}
  B_n(\lambda)=1-\frac{2\lambda}{(2n)^\alpha} +
  \frac{\lambda^2}{(2n)^{2\alpha}} +\widetilde{O}_I(n^{-1-\alpha})\,,
\quad\text{ as}\quad n\to\infty\,.
\end{equation*}
Clearly,
\begin{equation*}
  C_n(\lambda)=1-\frac{\alpha}{2n}+O(n^{-2})\,,\quad\text{ as}\quad n\to\infty\,.
\end{equation*}
Thus, in view of (\ref{eq:weights-diagonal}) and (\ref{complicated-case}),
one concludes that, as  $n\to\infty$,
\begin{equation}
\label{eq:F-n-expansion}
  F_n(\lambda)=2+\frac{2^{1-\alpha}}{c_1c_2}\lambda n^{-\alpha}-
\frac{2^{-2\alpha}}{c_1c_2}\lambda^2n^{-2\alpha}
-\frac{\alpha}{n}\left(1+\frac{1}{2c_1c_2}\right)
+\widetilde{O}_I(n^{-1-\alpha})\,.
\end{equation}

From the definition of $G_n(\lambda)$ found in
(\ref{eq:f-g-definition}), after substituting the expressions for for
$\{q_n\}_{n=1}^\infty$ and $\{b_n\}_{n=1}^\infty$ given by
(\ref{eq:weights-diagonal}), one can easily verify that
\begin{equation*}
  G_n(\lambda)=A_n(\lambda)\frac{(2n-1)^\alpha}{(2n)^\alpha}\,.
\end{equation*}
Thus
\begin{equation}
  \label{eq:G-n-expansion}
  G_n(\lambda)=1-\frac{\alpha}{n}+\widetilde{O}_I(n^{-1-\alpha})
\qquad\text{ as }\quad n\to\infty\,.
\end{equation}

\section{Asymptotics of $\beta_n(\lambda)$}
\label{sec:asymptotics-beta}
As in Appendix~\ref{sec:asymptotics-coef} we assume that the sequences $\{q_n\}_{n=1}^\infty$ and
$\{b_n\}_{n=1}^\infty$ are defined by (\ref{eq:weights-diagonal}) with
(\ref{eq:alpha-def}), (\ref{complicated-case}) and $I$ is
any bounded interval of $\mathbb{R}_+$. Thus the uniform asymptotic
expansions (\ref{eq:F-n-expansion}) and (\ref{eq:G-n-expansion}) hold
true. By simple algebraic manipulations, one has
\begin{equation}
  \label{eq:Fn-Fn-1}
  \begin{split}
  F_n(\lambda)F_{n-1}(\lambda)=&4+\frac{2^{3-\alpha}}{c_1c_2}\lambda
  n^{-\alpha} +
  \frac{2^{2-2\alpha}}{c_1c_2}\left(\frac{1}{c_1c_2}-1\right)\lambda^2n^{-2\alpha}
    \\ &- \frac{4\alpha}{n}\left(1+\frac{1}{2c_1c_2}\right)
-\frac{2^{2-3\alpha}}{(c_1c_2)^2}\lambda^3n^{-3\alpha}
+ \widetilde{O}_I(n^{-1-\alpha})\,,
  \end{split}
\end{equation}
as $n\to\infty$. On the basis of this result we find the asymptotic
formula of $4/(F_n(\lambda)F_{n-1}(\lambda))$. Thus we pass on to the
following expansion as $n\to\infty$
\begin{equation}
  \label{eq:4-over-Fn-Fn-1}
\begin{split}
  \frac{4}{F_n(\lambda)F_{n-1}(\lambda)}=&
1+\gamma_1(\lambda)n^{-\alpha}+\gamma_2(\lambda)n^{-2\alpha}
+\gamma_3(\lambda)n^{-1}\\ &+\gamma_4(\lambda)n^{-3\alpha}
  + \widetilde{O}_I(n^{-1-\alpha})\,,
\end{split}
\end{equation}
where $\gamma_1(\lambda)$, $\gamma_2(\lambda)$, $\gamma_3(\lambda)$,
and $\gamma_4(\lambda)$ can be calculated
explicitly by multiplying this last expression by $
F_n(\lambda)F_{n-1}(\lambda)$ and taking into account
(\ref{eq:Fn-Fn-1}). Thus, for (\ref{eq:4-over-Fn-Fn-1}) to be true, it
must be that
\begin{align*}
  \gamma_1(\lambda)&=-\frac{2^{1-\alpha}}{c_1c_2}\lambda\,,&\qquad
\gamma_2(\lambda)&=\frac{2^{-2\alpha}}{c_1c_2}
\left(\frac{3}{c_1c_2}+1\right)\lambda^2\,,\\
\gamma_3(\lambda)&=\alpha\left(1+\frac{1}{2c_1c_2}\right)\,,&\qquad
\gamma_4(\lambda)&=-\frac{2^{-3\alpha}}{(c_2c_2)^2}\left(3+\frac{4}{c_1c_2}\right)\lambda^3\,.
\end{align*}
It remains to multiply (\ref{eq:4-over-Fn-Fn-1}) by $G_n(\lambda)$, using
(\ref{eq:G-n-expansion}), and subtract $1$ to conclude that the uniform
asymptotic expansion as $n\to\infty$ of $\{\beta_n(\lambda)\}$, given
in (\ref{eq:beta-x-def}), is
\begin{equation}
  \label{eq:beta-asymptotics}
  \begin{split}
  \beta_n(\lambda)=&-\frac{2^{1-\alpha}}{c_1c_2}\lambda n^{-\alpha}+
\frac{2^{-2\alpha}}{c_1c_2}\left(\frac{3}{c_1c_2}+1\right)\lambda^2n^{-2\alpha}
+\frac{\alpha}{2c_1c_2}n^{-1}\\
&-\frac{2^{-3\alpha}}{(c_1c_2)^2}\left(3+\frac{4}{c_1c_2}\right)\lambda^3n^{-3a}
+\widetilde{O}_I(n^{-1-\alpha})\,.
\end{split}
\end{equation}

\begin{acknowledgments}
  We thank S. Simonov for valuable comments.
\end{acknowledgments}

\end{document}